\documentclass[fleqn,usenatbib,useAMS]{aa}
\usepackage{graphicx}	
\usepackage{amsmath}	
\usepackage{amssymb}	
\usepackage[T1]{fontenc}
\usepackage{url,lineno,microtype}
\usepackage{aas_macros}
\usepackage{txfonts}
\usepackage{tikz,xcolor}
\usepackage{hyperref}
\usepackage{subcaption}
\usepackage{physics}
\hypersetup{
        colorlinks = true,
        urlcolor = blue,
        linkcolor = blue,
        citecolor=blue
}

\usepackage{babel}   
\usepackage{soul}

\usepackage{ulem}
\usepackage{soul}
\newcommand{\prodimo}{\textsc{ProDiMo }}

\begin{document} 
\title{Energetic particles accelerated via turbulent magnetic reconnection in protoplanetary discs -- I. Ionisation rates}

   \author{Brunn, V.
          \inst{1}
          \and
          Pucci, F. \inst{2,3}
          \and
          Marcowith, A. \inst{4}
          \and
          Padovani, M. \inst{1}
          \and
          Rab, Ch. \inst{5,6}
         \and
          Sauty, Ch. \inst{4,7}
          }

   \institute{INAF-Osservatorio Astrofisico di Arcetri, Largo E. Fermi 5, 50125 Firenze, Italy\
              \email{valentin.brunn@inaf.it}
         \and SETI Institute, Mountain View, CA, USA
         \and IAPS, INAF, Via del Fosso del Cavaliere, 100, 00133 Roma RM, Italia
         \and 
         Laboratoire Univers et Particules de Montpellier, Université de Montpellier/CNRS, place E. Bataillon, cc072, 34095 Montpellier, France
         \and University Observatory, Faculty of Physics, Ludwig-Maximilians-Universität München, Scheinerstr. 1, 81679 Munich, Germany
        \and Max-Planck-Institut für extraterrestrische Physik, Giessenbachstrasse 1, 85748 Garching, Germany
        \and Laboratoire d’\'etude de l’Univers et des ph\'enom\`enes eXtr\^emes, Observatoire de Paris, Universit\'e PSL, CNRS, F-92190 Meudon, France
}

   \date{}

    \abstract 
   {Ionisation controls the chemistry, thermal balance, and magnetic coupling in protoplanetary discs. However, standard ionisation vectors such as stellar UV, X-rays, Galactic cosmic rays might not be efficient enough, as UV/X-rays are attenuated rapidly with depth, while Galactic cosmic rays are modulated. Turbulence-induced magnetic reconnection in disc atmospheric layers offers a physically motivated, in situ source of energetic particles (EPs) that has never been considered.}
   {We quantify the ionisation and heating produced by EPs accelerated by turbulent reconnection, identify where they dominate over X-rays and Galactic cosmic rays, and determine energetic thresholds for their relevance. We provide scalable diagnostics tied to the local energy budget.}
   {We adopt a Fermi-like acceleration model with parameters linked to a turbulent reconnection geometry trigger by the magneto-rotational instability, yielding a steady-state energy distribution of the EP forming a power-law of index $p=2.5$. We propagate electrons and protons through the disc and compute primary and secondary ionisation and associated heating on a fiducial T Tauri disc model background. The non-thermal normalisation is set by the fraction of local viscous accretion energy dissipation channelled to EPs, parametrised by $\kappa$.}
   {For $\kappa\gtrsim 0.4 \%$, EPs ionisation overpass standard sources such as X-rays and Galactic cosmic rays in the disc atmosphere and intermediate/deep layers out to radii of a few tens of astronomical units. Even at $\kappa\sim 0.025\%$, EPs contribute at the few-percent level, thus are chemically and dynamically relevant. The EP-induced heating complements UV/X-ray heating in the atmosphere and persists deeper. These results identify EPs accelerated by turbulence-induced magnetic reconnection as a rather robust, disc-internal ionisation channel that should be included in thermo-chemical and dynamical models of protoplanetary discs.}
  {}

   \keywords{ Protoplanetary discs --  Magnetic reconnection -- Ionisation}

\maketitle

\section{Introduction}\label{S:INTRO}
Young, low-mass pre-main-sequence stars in the T Tauri phase, typically a few million years old, continue to contract and to accrete material from their circumstellar discs. How they remove the excess angular momentum to enable accretion remains an open problem \citep{2014A&A...570A..82V,hartmann2016accretion,2016A&A...591L...3M}. Most viable scenarios invoke magnetic fields coupled to partly ionised gas, through either turbulent torques driven by magneto-hydrodynamic (MHD) instabilities \citep{1973A&A....24..337S}, or laminar torques associated with magnetised winds and jets \citep{blandford1982hydromagnetic,2019MNRAS.490.3112J}. In this work, we focus on the consequences of MHD turbulence, where present in the disc. We restrict our attention to protoplanetary discs where the magneto-rotational instability (MRI) is likely the primary turbulence driver
\citep{1991ApJ...376..214B,2003ARA&A..41..555B}. Crucially, in any accretion scenario where magnetic stresses regulate angular momentum transport, magnetic energy must be dissipated. In MRI-active layers, current sheets form so that dissipation through magnetic reconnection is a natural by-product of turbulent, magnetised discs \citep{2024ApJ...970...87P}.\\ 
Reconnection can occur in several young stellar object (YSO) environments: in stellar coronae during flares \citep{2017LRSP...14....2B}, in the star–disc interaction region near the truncation radius \citep{2000MNRAS.312..387F,2023ApJ...954...15F}, and, central to this paper, within the disc’s MRI-active layers, where field-line turbulence continually regenerates current sheets \citep{2010A&A...518A...5D}. Besides, sustaining MRI requires a minimum ionisation level. In YSO environments several sources of ionisation have already been investigated: UV and X-ray radiation either coming directly from the star or from the background radiation field \citep{1997ApJ...480..344G}, Galactic cosmic rays (GCRs), \citep{Cleeves13,Padovani18}, radionuclide ionisation \citep{2013ApJ...777...28C}, stellar energetic particles (SEP ; \citealt{Rab17,Rodgers-Lee17}) and in-situ accelerated particles, in shocks \citep{2016A&A...590A...8P,2019ApJ...883..121O} or by magnetic reconnection in flares \citep{2023MNRAS.519.5673B,2024MNRAS.530.3669B}. In the latter work, reconnection-accelerated EPs were shown to be an exceptionally efficient ionisation source, exceeding all others in their injection region (inner disc, $R \lesssim 1$ au), which motivated the present work.\\
These sources are necessary to maintain sufficient disc ionisation to trigger MRI. Here we consider an additional, internal channel: non-thermal or energetic particles (EPs hereafter) produced in situ by turbulence-induced reconnection and their impact on the disc’s ionisation balance.\\
As MHD turbulence and magnetic reconnection develop in accretion discs, a part of the stored magnetic energy is possibly released in the form of non-thermal particles. Magnetic reconnection sites are known to accelerate non-thermal particles (see reviews in different astrophysical or space plasmas contexts by \citealt{2020LRCA....6....1M, 2023SSRv..219...75O, 2024SSRv..220...43G}). Particles can be accelerated in a variety of processes: Fermi-I like acceleration associated with the electric field induced in gas inflows \citep{2005A&A...441..845D,  2012MNRAS.422.2474D}, or due to curved magnetic fields. Acceleration can be due to the betatron effect in increasing magnetic zones under the conservation of the adiabatic particle moment, in magnetic contracting islands \citep{2010ApJ...714..915O} or by direct electric field acceleration \citep{2023SSRv..219...75O}. Whatever the exact acceleration process, magnetic reconnection can be seen as a mechanism able to inject non-thermal particles into large scale turbulent flows as magnetic reconnection and turbulence are intimately inter-related. 

In this work, we focus on the effect of non-thermal particles locally accelerated by the turbulence-induced magnetic reconnection. Rather than commit to one microphysical route, we adopt an empirical, Fermi-like parametrisation of the non-thermal tail and quantify the resulting ionisation and heating, treating the acceleration physics as an effective source term tied to the local turbulent–reconnection environment. A work in preparation, paper II,  will address the consequences of this new ionisation source for the disc chemistry and chemical species abundances.\\

The paper is organised as follows. Sec. \ref{S:PHYS} outlines the disc thermo-chemical set-up and links it to turbulence and reconnection scalings. Sec. \ref{S:REC} introduces our EP injection and acceleration prescription as well as the transport set-up. Sec. \ref{Sec:Ionisation} presents ionisation (and associated heating) maps and identifies where EPs dominate over X-rays and GCRs. Sec. \ref{S:DIS} examines implications and limitations. Sec. \ref{sec:conclusions} summarises the main findings and outlines future directions.

\section{Physical framework: From thermochemistry to turbulence-driven reconnection}\label{S:PHYS}
\begin{figure*}[h!]
   \centering
   \includegraphics[width=\textwidth]{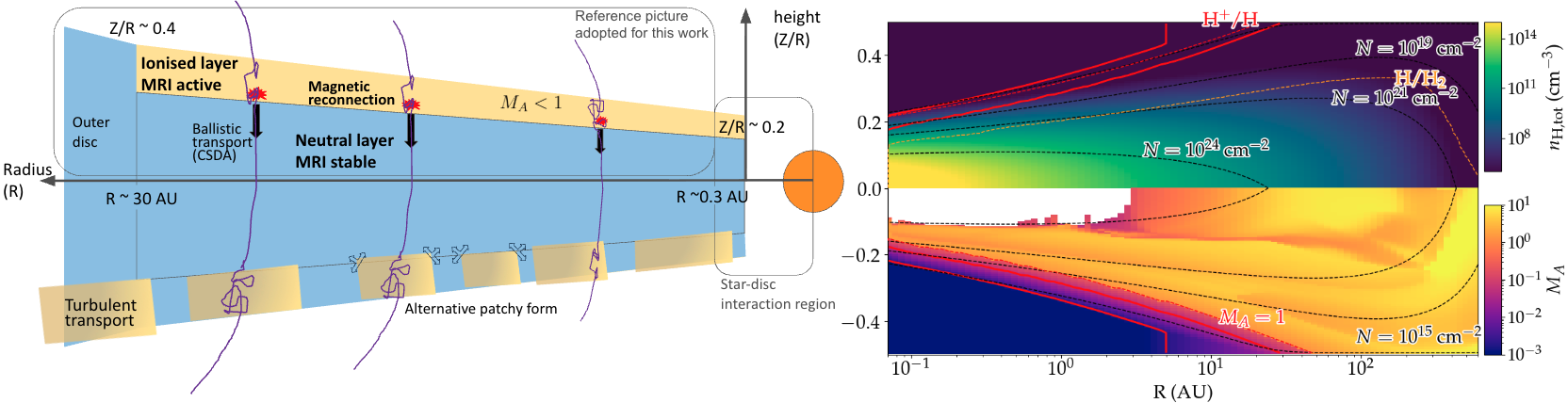}
   \caption{Left: Schematic radial cut from $R\sim0.1$ to $\sim 50$ AU presenting our scenario qualitatively. The disc is vertically stratified into a neutral mid-plane (blue; MRI-stable), where EPs arriving from above travel quasi-ballistically (CSDA-like) along near-vertical paths, and an ionised atmosphere (yellow; low-$\beta_p$ MRI-active), where sub-Alfv\'enic turbulence ($M_A<1$) triggers turbulent magnetic reconnection that injects EPs, which then stream downwards (black arrows). In the bottom left part, a patchy variant of the ionised atmospheric layer is sketched, with a lower covering area and possible lateral growth (black arrows). In the ionised turbulent layer, EP transport could be treated as stochastic. The inner star–disc interaction zone treated in \citet{2023MNRAS.519.5673B,2024MNRAS.530.3669B} is indicated at $R\lesssim0.3$. Right: Radial cut from $R\sim0.07$ to $\sim 600$ AU displaying the disc chemical and turbulent structure used in this work based on the \prodimo model. The top panel shows the total hydrogen number density, $n_{\rm H,tot}$ (colour), with vertical column-density contours $N=10^{15},10^{19},10^{21},10^{24}\,\mathrm{cm^{-2}}$ (dashed black). The dashed red line marks the ionised–atomic transition ($\mathrm{H^+/H}$) defining the boarder between the region where $\mathrm{H^+}$ dominates and the region where $\mathrm{H}$ dominates. Analogously, the dashed orange line delimits the atomic–molecular transition ($\mathrm{H/H_2}$). The bottom panel displays the Alfvénic Mach number $M_A\equiv v_{\rm turb}/V_A$ (colour). The red curve traces $M_A=1$. The sub-Alfvénic region ($M_A<1$) identifies where fast, turbulence-enabled reconnection, and thus EP acceleration, is expected. The solid red lines delimit the region where EP are efficiently accelerated, i.e. to energy more than 10 MeV. The white region is the so-called 'dead zone', where MRI is inefficient (see criterion Eq. \ref{eq:effectiveviscosityThi}), there is no turbulence, $M_A$ is set to 0.}
    \label{Fig:SKE}
    \end{figure*}

In this section we introduce the physical framework. We first outline the disc thermo-chemical background, then summarise observational and theoretical constraints on turbulence, and finally present the reconnection model that sets the parameters for EP acceleration. The left panel of Fig.~\ref{Fig:SKE} sketches our scenario: a protoplanetary disc with a weakly ionised, MRI-stable mid-plane overlain by a magnetically coupled atmosphere where sub-Alfvénic MRI turbulence bends field lines and builds current sheets. Magnetic reconnection in these atmospheric layers accelerates supra-thermal particles that propagate quasi-ballistically along near-vertical field lines \footnote{In this work we consider by default the case of vertical field lines transiting from the ionised layer towards the neutral disc. More complex magnetic geometries will be addressed in a forthcoming article (see e.g. \citealt{2023MNRAS.519.5673B} for a discussion of magnetic topologies in accretion discs).} towards the mid-plane. This section presents the quantitative arguments and assumptions that justify this scenario and defines the quantities used in the following model.

\subsection{Protoplanetary disc model (\prodimo) }
\textsc{ProDiMo}\footnote{\url{https://prodimo.iwf.oeaw.ac.at} \mbox{revision: 66efbd75 (2023/06/27)}} is a 2D radiation thermo-chemical code for protoplanetary discs \citep{woitke2009radiation,kamp2010radiation,thi2011radiation,woitke2016consistent}. It self-consistently couples disc structure, wavelength-dependent radiative transfer, gas and dust chemistry, and thermal balance to predict physical conditions and synthetic spectra over a broad wavelength range. The disc is treated on an axisymmetric grid (R,Z) for $R\approx 0.07 - 600$ AU and $Z/R= 0-0.5$ with a resolution set to $100\times100$. Frequency-dependent dust continuum radiative transfer yields the local radiation field and dust temperature, including stellar and interstellar UV and stellar X-rays, enabling photochemistry and heating and cooling in the atmosphere. The chemical network spans 235 species and 3143 reactions \citep{kamp2017consistent,Rab17}, with ice freeze-out and grain desorption \citep{thi2011radiation}. Heating processes include photoelectric emission, X-ray/cosmic-ray ionisation, viscous and collisional heating; cooling is treated via non-local thermal equilibrium line emission (e.g. CO, Fe II, O I), gas–dust coupling, and molecular lines. Iterative coupling ensures a consistent vertical structure \citep{2022A&A...666A.139A}.

We adopted the fiducial T Tauri disc of \citet{Rab17}, with $M_\star=0.7 M_\odot$, $T_\star=4000 \mathrm{K}$, $L_\star=1.0 L_\odot$, and an accretion rate of $\dot{M}=10^{-8} M_\odot \mathrm{yr}^{-1}$, as is detailed in their Table 4, for our thermo-chemical background. This set-up has been widely used \citep[][]{woitke2016consistent,kamp2017consistent,2019PASP..131f4301W}. From the model, we extracted at each grid cell $(R,Z)$: the total hydrogen nuclei density, $n_{\rm H,tot}=n_{\rm H}+2n_{\rm H_2}$; the gas temperature, $T$; the electron fraction, $x_e$; the local composition ($n_{\rm H}$, $n_{\rm H_2}$, $n_{\rm He}$) used to characterise the turbulence and to build mixture-averaged energy-loss functions for electrons and protons; and the vertical column densities, $N(R,Z)\equiv\int_{0.5R}^Z n_{\rm H,tot}(R,z) dz $, required for the transport model. These quantities set the injection EP energetics (Sect.\ref{S:REC}), determine energy losses during propagation (Sect.\ref{Sec:Ionisation}), and yield the reference X-ray and CR ionisation rates used for comparison.\\
The top right panel of Fig.~\ref{Fig:SKE} displays the disc structure used in this work. As per \prodimo, it shows a colour map of the total hydrogen number density, $n_{\mathrm{H,tot}}$. Dashed black contours trace the vertical column density, $N$, at $10^{15},\,10^{19},\,10^{21}$, and $10^{24}\,\mathrm{cm^{-2}}$. The dashed red curve marks the $\mathrm{H^+/H}$ transition (ionised–atomic boundary), and the dashed orange curve marks the $\mathrm{H/H_2}$ transition (atomic–molecular boundary). In the following, we adopt the following notation to designate the different regions of the disc: (i) Atmosphere: $10^{15} \le N < 10^{19}\,\mathrm{cm^{-2}}$, highly ionised.
(ii) Surface: $10^{19} \lesssim N \lesssim 10^{21}\,\mathrm{cm^{-2}}$, predominantly atomic gas.
(iii) Intermediate layer: $10^{21} \lesssim N\lesssim 10^{24}\,\mathrm{cm^{-2}}$, dominated by molecular gas.
(iv) Deep layer: $N \gtrsim 10^{24}\,\mathrm{cm^{-2}}$, characterised by strong visual extinction ($A_V\gtrsim10)$.

\subsection{Turbulence in protoplanetary discs}
In this section, we summarise the key observational constraints on the amplitude and vertical stratification of turbulence in protoplanetary discs, and outline the parametrisation we adopt to embed MRI-like turbulence into our static thermo-chemical disc model.

\subsubsection{Observational constraints on turbulence in discs} 
Quantifying turbulence in discs relies primarily on (i) spectrally resolved molecular line observations that separate thermal from non-thermal broadening, (ii) the use of multiple tracers and isotopologues that probe different heights, and (iii) indirect constraints from dust settling. High spectral--spatial resolution ALMA data of CO and its isotopologues generally indicate weak turbulence in many systems, particularly in the outer disc ($R\gtrsim 30$ AU) and near the mid-plane, with typical non-thermal velocities bounded by $\varv_{\rm turb} \lesssim 0.1 c_s$ at radii of a few  tens of astronomical units \citep[e.g.][]{2017ApJ...843..150F,2023NewAR..9601674R}. Independent inferences from the thickness of dust layers likewise point to low turbulent stirring. For instance, in HL~Tau, dust vertical stratification implies $\varv_{\rm turb} \sim 10^{-2} c_s$ at $R\sim 100$~AU \citep{2016ApJ...816...25P}. However, a contrasting picture emerges in some sources and at higher altitudes above the mid-plane. Broader CO lines in DM Tau suggest $\varv_{\rm turb} \sim 0.2\text{--}0.3 c_s$ at $R \sim 20\text{--}50$~AU \citep{2020ApJ...895..109F}. Moreover, chemically selective tracers that originate in the upper layers (e.g. CN, C$_2$H) reveal non-negligible turbulence at $Z/R \gtrsim 0.2\text{--}0.3$ and large radii in IM Lup \citep{2024A&A...684A.174P}. Recent ALMA CO analyses also in IM Lup, report on $\varv_{\rm turb} \sim 0.18\text{--}0.30 c_s$ at $R\sim 60$~AU \citep{2024MNRAS.532..363F}. Taken together, these measurements suggest a vertical gradient in the turbulent amplitude: turbulence is generally weakest near the mid-plane and stronger in the disc atmosphere, consistent with expectations for more ionised, MRI-active surface layers. This work focuses on a radial domain in the inner disc ($R \simeq 0.3$ to $30$~AU) that is not constrained by ALMA. The turbulence is unknown. However, it is expected that within this region MRI is active \citep{2023ASPC..534..465L}. The inner boundary reflects the zone of direct stellar magnetospheric influence due to stellar flaring loops, typically extending to $R\sim 0.3-0.5$~AU \citep{ 2024MNRAS.530.3669B}. 

\subsubsection{Turbulence modelling in a static disc model}
To evaluate the energy distribution of non-thermal particles, we first summarise the key properties of MHD turbulence in the region of interest, focusing on MRI-driven turbulence. MRI requires sufficient ionisation for strong gas–field coupling. The geometric set-up adopted here is sketched on the left part of Fig.~\ref{Fig:SKE}. 

Predicting the amplitude of turbulent velocity fluctuations remains challenging. The classical \citet{1973A&A....24..337S} framework parametrises angular-momentum transport via an effective viscosity,
\begin{equation}
    \nu_{\text{eff}}\equiv L \varv_{\rm turb} = \alpha_{\rm eff} c_s H,
    \label{eq:effectiveviscosity}
\end{equation}
where $L$ is a characteristic turbulent length scale, $\varv_{\rm turb}$ the turbulent speed, $c_s$ the sound speed, and $H$ the disc scale height. $\alpha_{\rm eff}<1$ encodes transport efficiency. \\
The velocity amplitude depends both on $\alpha_{\rm eff}$ and on how this efficiency partitions between spatial and velocity scales, as was discussed in \citet{2011ApJ...727...85H}. If eddies act on scales of $\sim H$, one expects $\varv_{\rm turb}\sim \alpha_{\rm eff} c_s$; if $\alpha_{\rm eff}$ distributes similarly across length and velocity, then $\varv_{\rm turb}\sim \sqrt{\alpha_{\rm eff}} c_s$. Empirically, observations and simulations rather support
\begin{eqnarray}
    \varv_{\rm turb}\approx \sqrt{\alpha_{\rm eff}} c_s
\quad \Rightarrow \quad
L \approx \sqrt{\alpha_{\rm eff}} H,
\label{eq:turbulentscale}
\end{eqnarray}
which we adopt as the representative eddy size \citep{2001ApJ...546..496C, 2011ApJ...727...85H}.\\
We evaluated $\alpha_{\rm eff}$ using the non-ideal, MRI-based prescription of \citet{2019A&A...632A..44T}, described in Appendix~\ref{annex:Viscosity}. This yields $\varv_{\rm turb}\approx 0.1\text{–}0.4 c_s$ in the upper layers and $\varv_{\rm turb}\lesssim 0.01 c_s$ near the mid-plane, consistent with observational constraints, and provides the values of the $L$ parameter encoding the reconnecting current sheet length and the turbulent speed used below in section \ref{S:MSCALE} to connect turbulence to reconnection geometry and, ultimately, to the non-thermal particle distribution.

\subsection{Magnetic reconnection and turbulence in accretion flows}
In this section, we show that the ionised, magnetised disc surface layers make MRI turbulence an efficient driver of current-sheet formation and magnetic reconnection, and we define the reconnection parameters that govern particle acceleration.

\subsubsection{Magnetohydrodynamic turbulence and magnetic reconnection}
Turbulence by itself does not guarantee magnetic reconnection. However, the atmospheric layers of protoplanetary discs combine physical conditions that make reconnection a natural outcome of the dynamics. Indeed, in this region, irradiation by stellar far UV and X-rays maintains high ionisation fractions ($\gtrsim 0.1$, see the H$^+$/H transition Fig. \ref{Fig:SKE}) ensuring tight magnetic coupling (weak Ohmic and ambipolar slippage). Where coupling is good, the MRI efficiently sustains turbulence and Maxwell stresses. The canonical expectation is that MRI is most efficient for high $\beta_p$ (>100); MRI in low-$\beta_p$ require additional conditions to grow. The parameter $\beta_p=P_{\text{gas}} / P_{\text{mag}}$ is the ratio of the gas pressure to the magnetic pressure. In particular, \citet{2000ApJ...540..372K} showed that reducing the toroidal component of the magnetic field re-opens the MRI window even at low $\beta_p$. Recent simulations of disc surface layers are consistent with this picture. Under conditions of strong ionisation and suitable field geometry, MRI-driven turbulence naturally generates magnetic reversals, shears and then thin current sheets, the precursors of reconnection \citep[e.g.][]{jacqueminetal2021,2021ApJ...920L..29R}, making this process frequent and energetically relevant in disc atmospheres.

There are only few works linking turbulence and magnetic reconnection in accretion discs. \citet{2018ApJ...864...52K} explore the connection between turbulence and fast magnetic reconnection in accreting systems. The authors conduct 3D ideal MHD simulations in the shearing box approximation. Although ideal, the simulations allow magnetic reconnection thanks to numerical resistivity. Their set-up starts from an initial azimuthal magnetic field and an exponential profile of the gas density over a scale $H$. The azimuthal magnetic field component dominates the initial radial and vertical components. The initial plasma beta parameter, $\beta_p$, is set to 1, 10 and 100 in a series of three different runs. The simulations follow the evolution of the magnetic components under the effect of the MRI but also of the Parker-Rayleigh-Taylor instability \citep[PRTI,][]{1966ApJ...145..811P}. 

The PRTI together with MRI, sustains an $\alpha-\Omega$ dynamo that reaches a quasi-steady state in which the characteristic reconnection speed is $V_r\approx(0.13\pm 0.09) V_A $ in the disc and $(0.17\pm 0.10) V_A$ in the corona, and the typical current-sheet width is $\Delta_r \approx 0.08 H$ (disc) and $0.10\, H$ (corona), where $V_A$ is the local Alfvén speed. We computed the scale height using the Keplerian thin disc relation $H=c_s/\Omega_K$. \footnote{Although the \textsc{ProDiMo} structure is moderately flared ($H/R\!\sim\!0.1$), the deviation of the angular frequency from Keplerian due to pressure support is $\mathcal{O}[(H/R)^2]\!\sim\!10^{-2}$ \citep{2019SAAS...45....1A}. This percent-level correction is well within typical astrophysical uncertainties, so the Keplerian approximation is adequate for our purposes.}

We report these numbers to show that a simulation, which self-consistently evolves MRI turbulence and reconnection, within the turbulence regime relevant to our work ($\varv_{\rm turb} \approx 0.1 - 0.4 c_s$), naturally results in fast magnetic reconnection. Moreover, by applying the static viscous framework adopted here \citep[see Appendix \ref{annex:Viscosity},][]{2019A&A...632A..44T} to our \textsc{ProDiMo} thermochemical background, we estimated the local Alfvén speed, $V_A = c_s \sqrt{\frac{2}{\beta_p}}$, and the Alfv\'enic Mach number, $M_A \equiv \frac{\varv_{\rm turb}}{V_A}$, where $c_s$ is the local sound speed and $\varv_{\rm turb}$ the turbulent velocity.  From this, we derived the reconnection speed, $V_r$, from \citet{1999ApJ...517..700L} scaling, a model particularly suitable in the context of turbulent protoplanetary disc regions. In sub- to trans-Alfv\'enic turbulence ($M_A\lesssim 1$), $V_r \sim V_A M_A^2$, we predict $V_r \sim 0.16 V_A$ at the disc surface at 1 AU from the star, in very good agreement with the \citet{2018ApJ...864...52K} results discussed above. 
The bottom right panel of Fig.~\ref{Fig:SKE} summarises the turbulent state we derived here and used to set the reconnection parameters on the \textsc{ProDiMo} grid. It shows a colour map of $M_A$, with the dashed red lines marking the $M_A=1$ contour. The low-$\beta_p$ ($\lesssim 1$) sub-Alfv\'enic domain ($M_A<1$), bounded below this line, is the region of interest for EP acceleration, as it is the regime in which fast, turbulence-enabled magnetic reconnection is expected. The white region is the so called 'dead zone', where MRI is inefficient, there is no turbulence, and $M_A \equiv 0$.\\
The values of \{$V_r, \Delta_r, H, V_A, M_A$\} computed from \prodimo thus provide a coherent, simulation-supported model from which we estimate the reconnection–driven EP acceleration parameters, as is detailed in Sect. \ref{sec:accelerationparameters}.

\subsubsection{Multi-scale magnetic reconnection modelling and particle acceleration}\label{S:MSCALE}

At scales larger than kinetic ones, magnetic reconnection is allowed through the breaking of magnetic frozen-in conditions by collisions, even when the plasma is fully ionised and neutrals are not present (see e.g. \citealt{Sweet:1958, 1963PhFl....6..459F}). 
While at kinetic scales magnetic reconnection is known to be sustained by electron inertia or the electron pressure tensor and to be efficient (i.e. fast) in converting magnetic energy into particle acceleration, fluid models have struggled to provide an explanation for fast energy release. Different models have been proposed in recent years (see e.g. the discussion in \citealt{2020PhPl...27a2305L}) , of which the tearing instability on thin current sheets is the most promising (see e.g. \citealt{Loureiro:2007,2009APS..DPPJM9003B,PucciVelli:2014}). The survival of laminar current sheets, necessary for the onset of the tearing instability, remains controversial in the context of turbulent protoplanetary discs, and outside the scope of this work. Still, reconnection is observed in 2D and also 3D turbulent simulations even with sufficiently high magnetic Reynolds number (see \citealt{doi:10.1126/sciadv.abn7627} and references therein). In the latter work, the formation of reconnection-produced magnetic flux ropes in 3D is shown to exhibit a complex morphology, differing significantly from the simpler island-like morphology in 2D. The authors note the similar size of the flux ropes between the 2D and 3D configurations, though the size of plasmoids is distributed over a wide variety of scales.

\begin{figure}[h!]
    \centering
    \includegraphics[width=0.8\linewidth]{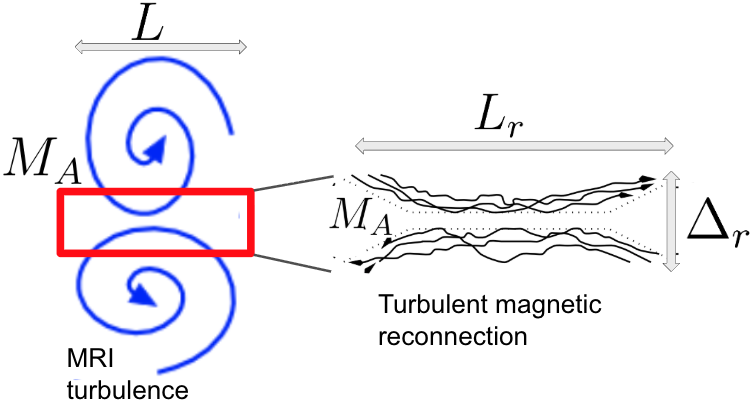}
    \caption{Sketch of MRI-driven turbulence and embedded turbulent magnetic reconnection. Left: Eddies at the injection scale $L$ in a sub-Alfv\'enic regime characterised by the Alfv\'enic Mach number $M_A$. Right: Reconnection layer produced by the turbulence, with length $L_r=L$ and thickness $\Delta_r$. The turbulence parameters \{$L,M_A$\} set the reconnection, geometry parameters \{$L_r,\Delta_r$\}, speed $V_{\rm rec}\sim V_A M_A^2$, and EP acceleration parameters used in this work.}
    \label{fig:MRIInducedMagneticReconnection}
\end{figure}

The length, $L_r$, of the largest current sheet forming in a turbulent region is assumed to be approximately equal to the local turbulent injection scale, i.e. $L \approx L_r$; the largest current sheets would be destroyed by the turbulence. The thickness, $\Delta_r$, is set by the scale over which a magnetic field line deviates from its original direction (see \citealt{1999ApJ...517..700L}) or, adopting a laminar tearing model, the intrinsic scale at which reconnection becomes fast \citep{2009APS..DPPJM9003B, Loureiro:2007, PucciVelli:2014}. Fig. \ref{fig:MRIInducedMagneticReconnection} schematically illustrates our working picture: sub-Alfvénic MRI turbulence ($M_A<1$) at the injection scale, $L$ (left), generates current sheets where turbulent magnetic reconnection operates (right). The reconnecting current sheet has a typical length, $L_r$, and thickness, $\Delta_r$, with $L_r = L$. 

In a generic magnetic reconnection geometry, EP acceleration is characterised by three parameters – the acceleration rate ,$\alpha_{\rm acc}$, the escape time, $\tau_{\rm esc}$, and the injection time, $\tau_{\rm inj}$ – which depend on the reconnection model and on the characteristic size and lifetime of the reconnecting current sheet \citep{1999ApJ...517..700L,2012PhRvL.108x1102K,2020PhPl...27a2305L,2023ApJ...942...21X},
\begin{equation}
    \tau_{\rm inj}=\frac{L_r}{V_r},\qquad \tau_{\rm esc}=\frac{L_r}{V_A}.
    \label{eq:definitiontauinjtauesc}
\end{equation}
For the purpose of this paper we consider the developments of the turbulent-reconnection framework of \citet{2023JHEAp..40....1Z} to be appropriate for the estimation of the reconnection parameters. Thus, the turbulent parameters \{$L,M_A$\} set the reconnection speed, $V_{\rm rec} \sim V_A M_A^2$ \citep{1999ApJ...517..700L}, and, as is shown below, the parameters of the EP acceleration model. The particles are considered to be accelerated multiple times within the turbulent layer, transitioning between different current sheets, and undergo Fermi acceleration through island compression.

The same assumption holds when assuming that the plasmoid instability or any fast tearing instability kicks in. In particular we consider here the 'ideal' tearing instability (see \citealt{PucciVelli:2014}). 
Then the acceleration rate is $\alpha_{\rm acc}\sim t^{-1}_A$, where the proportionality constant depends on the topology of the reconnecting magnetic field \citep{2018PhPl...25c2113P} and other non-ideal effects (see e.g. \citealt{2017ApJ...845...25P,2019ApJ...883..172S}). 
In this work, we adopt the results of \citet{2023ApJ...942...21X}, giving
\begin{equation}
    \alpha_{\rm acc}=\frac{2V_r}{3\Delta_r},\qquad \Delta_r=L_r M_A^2,\qquad V_r=V_A M_A^2,
    \label{eq:accelerationrate}
\end{equation}
and together with Eq. \eqref{eq:definitiontauinjtauesc}, 
\begin{equation}
    \alpha_{\rm acc}=\frac{2}{3} t_A^{-1},\qquad
\tau_{\rm inj}=M_A^{-2} t_A,\qquad
\tau_{\rm esc}=t_A,
\label{eq:reconnectionparameters}
\end{equation}
where $t_A\equiv L_r/V_A=\Delta_r/V_r$. 

In conclusion, comparing collisional fluid modelling to \citet{2023JHEAp..40....1Z} brings correction factors to the energy effectively converted within the turbulent layer in the disc's atmosphere. The development of detailed particle acceleration models based on the interplay of reconnection and turbulence remains largely phenomenological, and various types of dimensional analysis may be adopted. Not knowing the details of the microscopic acceleration process at this stage, we introduce the efficiency parameter, $\kappa$, and defer a full treatment to future work. 

\section{Methodology -- Energetic particle acceleration by turbulence-induced magnetic reconnection} \label{S:REC}
Low-energy ($\lesssim1$ GeV) particles can significantly enhance the ionisation rate in the disc, influencing both its chemistry and the coupling between the gas and magnetic fields \citep{Rab17,Padovani18, 2023MNRAS.519.5673B, 2024MNRAS.530.3669B}. In order to study ionisation by non-thermal particles in protoplanetary discs, we start by building a model to determine the energy distribution of the supra-thermal particles produced during magnetic reconnection events. 

\subsection{Energy distribution from acceleration by a Fermi-like process}
\label{sec:accelerationprocess}
Magnetic reconnection can efficiently accelerate a fraction of the thermal particle population to supra-thermal energies through Fermi-like acceleration mechanism. Assuming such an acceleration process is active in the region of interest, the energy gain by a particle is proportional to the particle kinetic energy, $\dot{E}=\alpha_{\rm acc} E$, with $\alpha_{\rm acc}$ being the acceleration rate. The equation for the energy distribution is \citep{2014PhRvL.113o5005G, 2015ApJ...806..167G,2023SSRv..219...75O}
\begin{equation}
    \frac{\partial F_{\rm nt}}{\partial t } + \frac{\partial}{\partial \epsilon}(\dot{E}F_{\rm nt} )= \frac{F_{\rm inj}}{\tau_{\rm inj}} - \frac{F_{\rm nt}}{\tau_{\rm esc}},
    \label{eq:fermiaccdistributioneq}
\end{equation}
where $F_{\rm nt}$ is the particle energy distribution (erg$^{-1}$ cm$^{-3}$), $\dot{E}$ is the particle energy gain rate (erg s$^{-1}$), $\tau_{\rm inj}$ is the timescale for injection of the initial particle distribution, $F_{\rm inj}$, into the accelerating region, and $\tau_{\rm esc}$ is the escape timescale from the acceleration region.\\ 
Assuming that the initial upstream distribution is a Maxwellian distribution \footnote{The analysis performed in \citet{2014PhRvL.113o5005G} has been developed in the context of plasmas at high magnetisation, but the general form of the solution does not depend on the magnetisation regime, it simply assumes Fermi-I type acceleration, an injection and an escape rate. The solution in Eq. \ref{eq:nonthermaldistribution} assumes non-relativistic temperature of the background plasma, which is the case in the astrophysical system under consideration.},
\begin{equation}
    F_{\rm inj}(\epsilon)=\frac{2 n_{\rm inj}}{\sqrt{\pi}}  \sqrt{\epsilon} \exp(-\epsilon),
    \label{eq:thermaldistribution}
\end{equation}
where $F_{\rm inj}$ is the upstream particle energy distribution in units of thermal energy, $\epsilon$ is the energy in units of thermal energy, $\epsilon=E/E_{\rm th}$, with $E_{\rm th}=3/2 k_B T$, where $k_B$ is the Boltzmann constant and $T$ the temperature of the medium, and $n_{\rm inj}$ is the density of particles entering in the acceleration region.\\
The solution of Eq. \eqref{eq:fermiaccdistributioneq} is a non-thermal distribution that can be written in units of thermal energy as
\begin{equation}
    F_{\rm nt}(\epsilon,t)= \frac{2 n_{\rm inj}}{\sqrt{\pi} \alpha_{\rm acc} \tau_{\rm inj}} \epsilon^{-(1+\beta)}\left[\Gamma_{3/2+\beta}(\epsilon e^{-\alpha_{\rm acc} t})-\Gamma_{3/2+\beta}(\epsilon)\right],
    \label{eq:nonthermaldistribution}
\end{equation}
where $\beta = \frac{1}{\alpha_{\rm acc} \tau_{\rm esc}}$ and $\Gamma(3/2+\beta,\epsilon)$ is the upper incomplete Gamma function of order $3/2+\beta$.\\
In the next sections, we evaluate the acceleration parameters assuming a specific turbulent magnetic reconnection model and see how to normalise it based on the power released by the turbulence in the disc. 

\subsection{Energy distribution parameters in turbulent magnetic reconnection} 
\label{sec:accelerationparameters}

The shape of the non-thermal energy distribution is set by the three parameters, $\alpha_{\rm acc}$, $\tau_{\rm esc}$, and $\tau_{\rm inj}$, as defined in Eq. \eqref{eq:reconnectionparameters}, which depend on the characteristic size and lifetime of the reconnecting current sheet.\\ 
As reconnection proceeds, particles are continuously injected and accelerated within current sheets. The combination of sustained injection and Fermi-like energisation naturally builds a supra-thermal tail. We take the effective acceleration duration to be the stability time of a reconnecting current sheet, $\tau_{\rm CS}$, which in a turbulent medium we approximate by the eddy turnover time at the injection scale,
$\tau_{\rm CS}=\frac{L_r}{\varv_{\rm turb}}$.
In what follows, we evaluate the non-thermal distribution at $t=\tau_{\rm CS}$. This 'end-of-burst' snapshot captures the particle population just before the sheet is disrupted, provides a representative reference for intermittent reconnection in turbulence, and simplifies the analysis for this first, stationary treatment.\\
After a time, $\tau_{\rm CS}$, a power-law component emerges if the factor $\Gamma_{3/2+\beta} \left(\epsilon e^{-\alpha_{\rm acc}\tau_{\rm CS}}\right)-\Gamma_{3/2+\beta}(\epsilon)$ in Eq.~\eqref{eq:nonthermaldistribution} is (approximately) energy-independent. This holds across the supra-thermal range provided $\epsilon e^{-\alpha_{\rm acc}\tau_{\rm CS}} \lesssim 1$. For $\epsilon e^{-\alpha_{\rm acc}\tau_{\rm CS}} \gtrsim 1$, the spectrum bends away from a pure power law, defining a finite-time cut-off,
\begin{equation}
    \epsilon_{\max,{\rm CS}}=\exp \left(\alpha_{\rm acc}\tau_{\rm CS}\right)=\exp \left(\frac{2}{3M_A}\right),
    \label{eq:energymaxcs}
\end{equation}
which directly reflects Fermi-like energisation, with $\dot\epsilon=\alpha_{\rm acc}\epsilon$ acting for a duration, $\tau_{\rm CS}$.\\
A second limit comes from confinement (Hillas criterion): the Larmor radius must fit within the accelerator, giving
\begin{equation}
    \epsilon_{\max,{\rm H}}=\frac{e B \Delta_r}{E_{\rm th}},
\end{equation}
with $B$ the magnetic field strength. The effective maximum energy is therefore
\begin{equation}
    E_{\max}=\min\{\epsilon_{\max,{\rm CS}},\ \epsilon_{\max,{\rm H}}\} E_{\rm th}.
\end{equation}
At the low-energy end, we define the injection threshold, $\epsilon_{\rm inj}$, by continuity between thermal and non-thermal components,
\begin{equation}
\label{eq:injectionenergyconstrain}
F_{\rm inj}(\epsilon_{\rm inj})=F_{\rm nt}(\epsilon_{\rm inj})\,,
\end{equation}
which we solve numerically at time $t=\tau_{CS}$. We then refer to the non-thermal population as EPs with normalised energy $\epsilon\ge\epsilon_{\rm inj}$.\\
The slope of the supra-thermal tail follows directly from the acceleration and escape parameters,
\begin{equation}
    p=1+\frac{1}{\alpha_{\rm acc}\tau_{\rm esc}}=2.5\,.
\end{equation}
In this work we take $\alpha_{\rm acc}$, $\tau_{\rm esc}$, and $\tau_{\rm inj}$ to be the same for electrons and protons, so the derived $p=2.5$ applies to both species. A species-dependent treatment is deferred to future work.\\
With the spectral shape fixed by $\alpha_{\rm acc}$, $\tau_{\rm inj}$, and $\tau_{\rm esc}$ (thus setting $p$, $\epsilon_{\rm inj}$, and $\epsilon_{\max}$), the remaining task is to determine the normalisation, i.e. the number density of non-thermal particles, from the local energy budget. In the next subsection, we link the particle power to a fraction of the turbulent accretion dissipation to obtain the non-thermal EP density.

\subsection{Normalisation of the energy distribution from the available energy of the turbulent accretion}
\label{sec:energetic constrains}

We normalised the non-thermal distribution by assuming that the power sustaining it, $\dot U_{\rm nt}$, is a fixed fraction, $\kappa$, of the local viscous dissipation per unit volume, $D/H$,
\begin{equation}
\dot U_{\rm nt}\equiv \int_{E_{\rm inj}}^{E_{\rm max}} F(E)\dot E{\rm d}E = \kappa\frac{D}{H},
\label{eq:hypothesisviscacc}
\end{equation}
where $D$ is the viscous energy flux (erg s$^{-1}$ cm$^{-2}$), $H$ the disc scale height, and $\kappa$ the (dimensionless) fraction channelled into non-thermal particle acceleration.\\
As we assumed a Fermi-like acceleration process, $\dot{E}=\alpha_{\rm acc} E$, with  $\alpha_{\rm acc}$ independent of energy, and thus
\begin{equation}
     \int_{E_{\rm inj}}^{E_{\rm max}} F(E) E \dd E = \kappa \frac{D}{\alpha_{\rm acc} H} ,
     \label{eq:energydensity}
\end{equation}
so the left-hand side is the non-thermal energy density, $U_{\rm nt}$, by definition.\\
In the Keplerian, thin disc approximation, the viscous dissipation is \citep{1981ARA&A..19..137P,2019NewA...70....7M}
\begin{equation}
D(R)=\frac{1}{2} \nu_{\rm eff}\Sigma\left(R\frac{{\rm d}\Omega}{{\rm d}R}\right)^2
=\frac{9}{8}\nu_{\rm eff}\Sigma\Omega^2,
\end{equation}
where $\nu_{\rm eff}$ is defined in Eq.~\eqref{eq:effectiveviscosity}, $\Sigma=H\rho$ is the accreting column density, $\rho$ is the mass density, $\Omega$ is the angular velocity, and $R$ is the distance to the star. Hence,
\begin{equation}
\frac{D}{H}
=\frac{9}{8} \rho c_s^2 \frac{L_r \varv_{\rm turb}}{H^2}
=\frac{5}{4} U_{\rm th} \frac{L_r \varv_{\rm turb}}{H^2},
\end{equation}
where we used $c_s=H\Omega$ and $U_{\rm th}=\tfrac{3}{2}P_{\rm th}$ with $\tfrac{5}{3} P_{\rm th} = \rho c_s^2$, as the disc is composed of mono-atomic gas in the acceleration region. Substituting into Eq.\eqref{eq:energydensity} gives
\begin{equation}
\frac{U_{\rm nt}}{U_{\rm th}} = \kappa\frac{5}{4\alpha_{\rm acc}}\frac{L_r \varv_{\rm turb}}{H^2}.
\end{equation}

\begin{figure}
    \centering
    \includegraphics[width=0.9\linewidth]{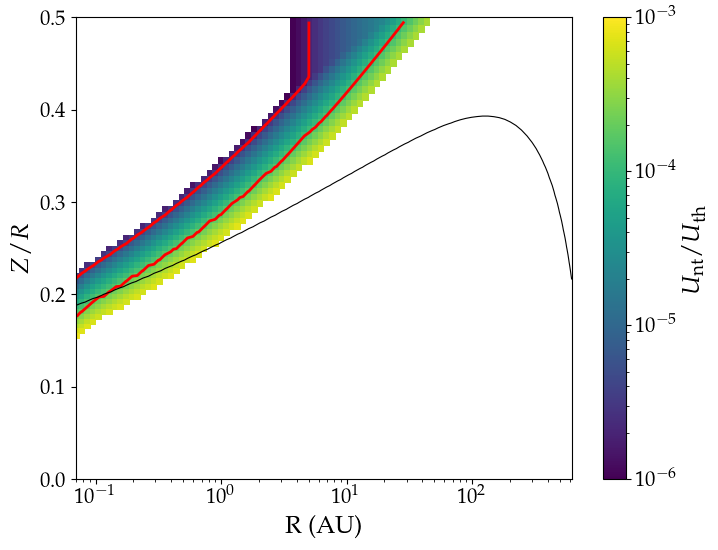}
    \caption{Non-thermal–to–thermal energy density ratio, $U_{\rm nt}/U_{\rm th}$, for $\kappa=0.1$. The red contour encloses regions where the local maximum particle energy exceeds $10~\mathrm{MeV}$ (efficient acceleration sites capable of supplying ionising EPs). The black contour marks a vertical column density, $N=10^{19} \mathrm{cm^{-2}}$ (approximate disc surface).}
    \label{fig:EnergyDensity}
\end{figure}

This condition fixes the normalisation. Hence the injected number density, $n_{\rm inj}$, of the distribution of Eq.\eqref{eq:nonthermaldistribution} is
\begin{equation}
n_{\rm inj}\gamma = \kappa \frac{5 \sqrt{\pi}}{8} \frac{L_r \tau_{\rm inj} \varv_{\rm turb}}{H^2} n_{\rm th},
\end{equation}
where $\gamma=\int_{\epsilon_{\rm inj}}^{\epsilon_{\rm max}} \epsilon^{-\beta}\left[\Gamma_{3/2+\beta}(\epsilon /\epsilon_{\rm max})-\Gamma_{3/2+\beta}(\epsilon)\right] d\epsilon$ and $n_{\rm th} = U_{\rm th}/E_{\rm th}$.
Using $\tau_{\rm inj} \varv_{\rm turb}=L_r/M_A$ (Eq.\ref{eq:reconnectionparameters}) yields
\begin{eqnarray}
    \frac{n_{\rm inj}}{n_{\rm th}}&=& \kappa \frac{5 \sqrt{\pi}}{8} \gamma^{-1} \left(\frac{L_r}{H}\right)^2 M_A^{-1} \,,
    \label{eq:ntnumberdensity}
\end{eqnarray}
where $L_r/H$ follows Eq.\eqref{eq:turbulentscale}, $\beta=1/(\alpha_{\rm acc}\tau_{\rm esc})=1.5$ from Eq.\eqref{eq:reconnectionparameters}, $M_A$ is computed locally (Appendix~\ref{annex:Viscosity}), $\epsilon_{\rm inj}$ is obtained by numerically solving Eq.~\eqref{eq:injectionenergyconstrain}, $\gamma$ is likewise solved numerically, and $\kappa$ remains the free energetics parameter.\\
Using the injected non-thermal density from Eq.\eqref{eq:ntnumberdensity}, we computed the local non-thermal energy density, $U_{\rm nt}(R,Z)$. Because the thermochemical background, and thus the acceleration parameters, vary with the position, all quantities were evaluated cell by cell on the \textsc{ProDiMo} (R,Z) grid. Fig. \ref{fig:EnergyDensity} maps the resulting $U_{\rm nt}/U_{\rm th}$ and marks, in red, the regions where the local maximum particle energy exceeds $10 \mathrm{MeV}$ (efficient acceleration \footnote{By efficient we mean, particles with energies above the peak of ionisation cross section.}), while the black contour traces $N=10^{19} \mathrm{cm^{-2}}$ (the disc surface). We set the non-thermal power to $\kappa = 0.1$, by analogy with the canonical $\sim 10 - 30\%$ efficiency with which supernova-remnant shocks are thought to channel kinetic energy into GCR \citep{2010ApJ...718...31P}. \\
On Fig. \ref{fig:EnergyDensity}, the sharp drop from $U_{\rm nt}>0$ to $U_{\rm nt}=0$ on the disc side tracks the $M_A=1$ contour. 
In our prescription, the maximum non-thermal energy $\varepsilon_{\max}$ tends to unity for $M_A \gtrsim 1$ (Eq.\eqref{eq:energymaxcs}), the supra-thermal tail merges with the thermal distribution. We therefore set $U_{\mathrm{nt}}=0$ for $M_A \gtrsim 1$, noting that this reflects the disappearance of a distinct non-thermal particle population rather than a suppression of turbulent reconnection itself. Moreover, because the turbulent reconnection model is formulated for strongly magnetised, low-$\beta_p$ plasmas \citep{1999ApJ...517..700L}, and simulations indicate that as $\beta_p$ increases, the non-thermal particle acceleration efficiency declines \citep{2017ApJ...843...21L}. We restricted EP acceleration to low-$\beta_p$ regions and set $U_{\mathrm{nt}}=0$ for $\beta>10$. We note that efficient non-thermal acceleration may occur in hot or relativistic high-$\beta$ plasmas, yet such regimes are not relevant for protoplanetary discs, and therefore lie outside the scope of this work.\\
On the atmosphere side, at very low $M_A$, $U_{\rm nt}$ also vanishes because the reconnection layer thickness scales as $\Delta_r \sim  M_A^2$ (Eq. \ref{eq:accelerationrate}), becoming extremely thin. The maximum particle energy is then limited by the Hillas criterion. In practice, we set $U_{\rm nt}=0$ when $E_{\max}\!\lesssim\!10 E_{\rm th}$.\\
Notably, the map shows that our turbulence-driven reconnection model can accelerate particles to mega-electronvolt energies in extended disc region, out to a few tens of astronomical units along the ionised atmospheric layers, i.e. within the intermediate window where $M_A<1$ yet not so small that $\Delta$ allow $E_{\max}\gg E_{\rm th}$.

Throughout the acceleration region, $U_{\rm nt}$ is orders of magnitude smaller than the thermal energy density, $U_{th}$, and, since $\beta_p\lesssim 1$, also far below the magnetic energy density. Consequently, EPs are not expected to exert a dynamical back-reaction over the plasma. This validates our treatment in the next section where EP transport is computed in the test-particle limit as a first approximation.
In the next section (i.e. Sect. \ref{subsec:nt_injection_propagation}), the non-thermal energy distribution is computed from this normalisation, together with the acceleration parameters at each \prodimo grid cell in order to model its propagation through the disc.

\section{Results: Ionisation from reconnection-accelerated particles}\label{Sec:Ionisation}

To quantify the ionisation rate in the disc due to EPs generated by turbulence induced magnetic reconnection, we first computed their initial energy distribution, propagation and energy attenuation through the disc material.

\subsection{Non-thermal particle energy distributions: injection and propagation}
\label{subsec:nt_injection_propagation}

\begin{figure}
    \centering
    \includegraphics[width=0.9\linewidth]{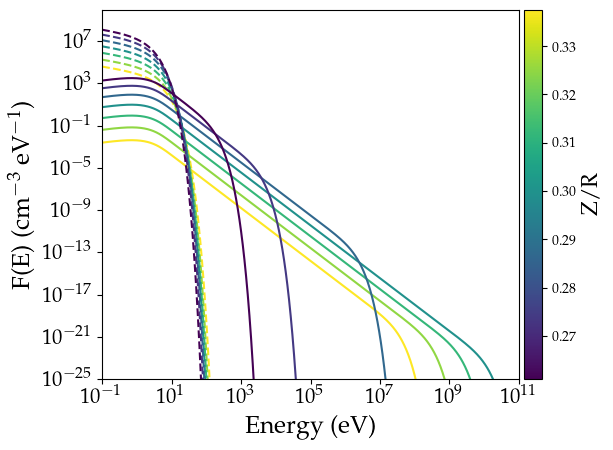}
    \caption{Particle energy distributions at $R=1$ AU for $\kappa=0.1$. The dashed curves show the thermal Maxwellian based on \textsc{ProDiMo} density and temperature for $Z/R=0.25\text{–}0.35$. The solid curves show the non-thermal distribution produced by magnetic-reconnection acceleration.}
    \label{fig:EPDistribution}
\end{figure}

For each \prodimo--{R,Z} grid cell, the local initial ('injection') non-thermal energy distribution $F^{R,Z}_{\rm nt}(E)$, was computed with Eq. \eqref{eq:nonthermaldistribution} at time $t=\tau_{\rm CS}$, using the normalisation of Eq. \eqref{eq:ntnumberdensity}, together with the acceleration parameters $\alpha_{\rm acc}$, $\tau_{\rm esc}$, and $\tau_{\rm inj}$. Fig.~\ref{fig:EPDistribution} shows representative energy distributions at $R=1~\mathrm{AU}$ within the height range where $U_{\rm nt}>0$: the thermal Maxwellian (dashed) based on \prodimo density and temperature, and the associated non-thermal tail (solid) produced by our model of magnetic reconnection in turbulence.
We then propagated this local energy distribution through the disc material with the continuous–slowing–down approximation (CSDA) to obtain, for any traversed column density, $N$, the propagated distributions, $F^{R,Z}_{\rm nt}(E,N)$, for both electrons and protons, as in \citet{2023MNRAS.519.5673B}.
We assume isotropic injection and approximate trajectories as vertical, which is adequate to first order in the subsonic–to–transonic turbulent regime considered here. A more complete treatment that explicitly includes turbulence, accounting for spatial diffusion along turbulent field lines and for stochastic re-acceleration in MRI turbulence \citep[e.g.][]{2021MNRAS.506.1128S}, is deferred to a forthcoming paper. Methodological details on loss functions, cross sections, CSDA validity, and the treatment of primaries and secondaries are provided by \citet{Padovani2009,Padovani18} and \citet{2023MNRAS.519.5673B}.

\subsection{Ionisation rate}
Energetic electrons and protons, produced in the turbulence-induced magnetic reconnection, propagate into the protoplanetary disc and interact with the ambient gas, leading to the ionisation of hydrogen and molecular hydrogen. To compute the ionisation rate, we consider both primary ionisation reactions and secondary processes\footnote{There is a typographical error in the expression of the differential ionisation cross section for electron in Eq. 19 of \citet{kim2000extension}. The right-hand side of the equation corresponds to $d\sigma/dw$, not $d\sigma/dW$. Confusing these two variables can introduce an error of a factor 13.6 in the differential cross section.} triggered by the ejected energetic electrons. The ionisation rate, $\zeta_{R,Z}(N)$, at a vertical column density, $N$, produced by the EP distribution, $F^{R,Z}_{\rm nt}(E)$, originating from a pixel at position $R,Z$, was computed as in \citet{2023MNRAS.519.5673B}.

The total EP-induced ionisation rate at radius $R$, $\zeta_R(N)$, was computed by vertically averaging the contributions from all reconnection-active pixels (cells with non-zero non-thermal energy density; the coloured pixels in Fig.~\ref{fig:EnergyDensity}). Each acceleration site at height $Z_i$ within the radial slice contributes by $\zeta_{R,Z_i}(N)$ to the total propagated ionisation profile. The total ionisation is

\begin{equation}
    \zeta_{\text{R}}(N) = \frac{1}{\Delta Z} \sum_i \zeta_{R,Z_i}(N)   \Delta Z_i\,,
    \label{eq:totalionisationrate}
\end{equation}
where $\Delta Z_i$ is the thickness of pixel $i$ and $\Delta Z=\sum_i \Delta Z_i$ is the total vertical extent of the active layer at that radius (i.e the length of the coloured slice at radius $R$ of Fig. \ref{fig:EnergyDensity}). 

We adopt a thickness-weighted vertical average because, at a fixed radius, multiple acceleration sites at different heights contribute simultaneously, and not equally, to the ionisation at depth $N$. Weighting each pixel’s contribution, $\zeta_{R,Z_i}(N)$, by its vertical extent, $\Delta Z_i$, and normalising by the total active thickness, $\Delta Z$, approximates the continuous integral, $\Delta Z^{-1} \int \zeta_{R,Z}(N) dZ$. This ensures that layers occupying a larger fraction of the reconnecting zone contribute proportionally more, eliminates biases due to grid resolution, and provides a surrogate for the time and ensemble average of inherently patchy, intermittent reconnection. We ignored EP ionisation from the opposite disc side because the propagated flux is suppressed before reaching the mid-plane at column density $N\gtrsim10^{25}\,\mathrm{cm^{-2}}$.

\begin{figure}
    \includegraphics[width=0.9\linewidth]{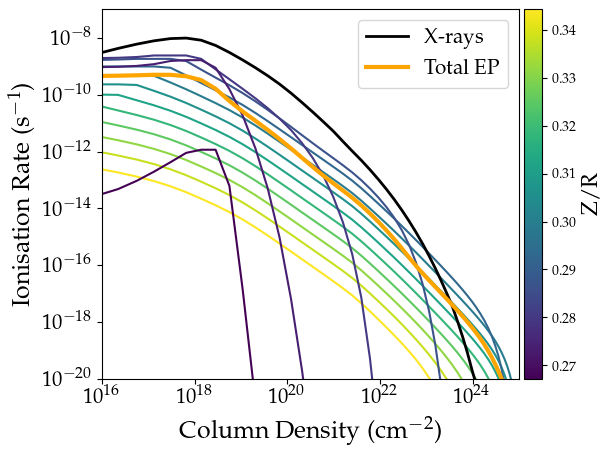}
    \caption{Ionisation rates as function of the disc vertical column density at $R=1$ AU for $\kappa=0.1$. The lines in the viridis colour scale are the contribution to the ionisation rate coming from EPs accelerated in region at different $Z/R$. The thick orange line is the total ionisation rate corresponding to the weighted sum (Eq. \ref{eq:totalionisationrate}) of each local contribution. The black line is the ionisation rate from the stellar X-rays. }
    \label{fig:IonisationRate1au}
\end{figure}

Fig. ~\ref{fig:IonisationRate1au} illustrates the result at $R=1 ~\mathrm{AU}$, for $\kappa=0.1$. Coloured curves (shown with the 'viridis' colour scale) are the individual $\zeta_{R,Z_i}(N)$ for EPs injected at different heights, the thick orange curve is the vertically averaged total from Eq.~\eqref{eq:totalionisationrate}, and the black curve shows, is for comparison, the stellar X-ray ionisation of a typical T Tauri object with an X-ray luminosity of $L_X=10^{30}$ erg s$^{-1}$. While X-rays dominate at low column density ($10^{16} - 10^{23} \mathrm{cm^{-2}}$), EPs substantially enhance and exceed the ionisation at higher columns (above a few $10^{23} \mathrm{cm^{-2}}$). At column density $N\gtrsim10^{25}\,\mathrm{cm^{-2}}$ (not shown in Fig. ~\ref{fig:IonisationRate1au}), the EP ionisation rate drops, and the GCRs component overtakes close to the disc mid-plain. The column density at which EPs, X-rays or GCR dominate ionisation depends on the distance to the star; we discuss this aspect in the next section.

\subsection{Spatial ionisation rate distribution}
\begin{figure}
    \centering
    \includegraphics[width=0.9\linewidth]{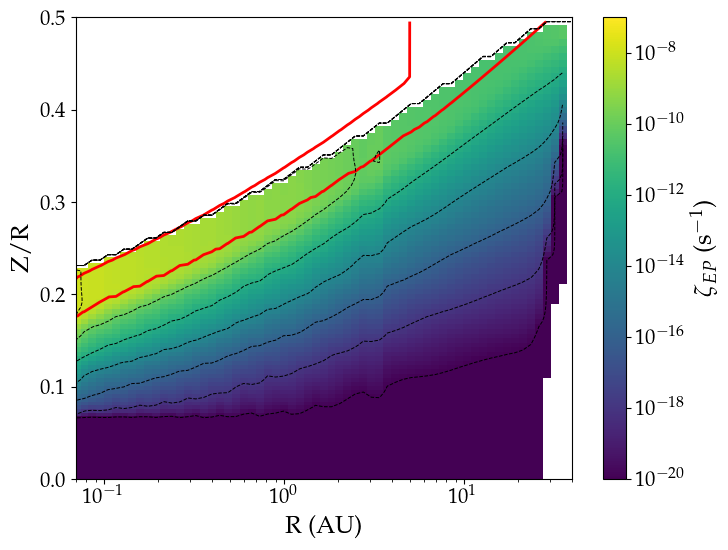}
    \caption{EP ionisation rate map for $\kappa=0.1$. The solid red contour outlines the region where EPs are accelerated. Dashed lines indicate isocontours of the EP ionisation rate. The upper white region corresponds to column densities below $N = 10^{15} \text{cm}^{-2}$, indicating regions above the disc atmosphere.}
    \label{fig:EPIonisationMap}
\end{figure}
Using Eq. \eqref{eq:totalionisationrate}, we computed the total EP–induced ionisation rate at every \prodimo grid cell $(R,Z)$, yielding the map in Fig. \ref{fig:EPIonisationMap}. The figure shows $\zeta_{\rm EP}$ on a logarithmic colour scale from $10^{-20}$s$^{-1}$ (dark purple) to $10^{-8} \mathrm{s^{-1}}$ (bright yellow). Dashed black curves are isocontours.

Ionisation is strongest in the inner disc and upper layers, then weakens with increasing radius and approaching the mid-plane, reflecting both attenuation with column density and reduced acceleration efficiency farther from the star. The solid red contour marks where particles are accelerated above $10~\mathrm{MeV}$, i.e. the portion of the atmosphere that efficiently injects EPs. The white band at the top indicates very low column density ($<10^{15} \mathrm{cm^{-2}}$) above the disc atmosphere. Outflow propagation in that region is beyond our present scope. The white area beyond $\sim 30$ AU corresponds to radii where acceleration is too weak to produce $>10~\mathrm{MeV}$ particles, so EPs are stopped at low column density and the ionisation is negligible there.

Importantly, while $\zeta_{\rm EP}$ peaks at the surface, EPs also penetrate to large column densities, sustaining significant ionisation ($N\gtrsim10^{23} \mathrm{cm^{-2}}$) where stellar X-rays are heavily attenuated. This deep reach positions EPs as the dominant ionisation source in otherwise shielded disc layers.

\subsubsection{EP-dominated ionisation regions}
This section identifies the conditions and locations where EPs dominate over X-rays and GCRs ionisation, mapping EP-dominated domains for $\kappa=0.1$ and 1. In line with \citet{Rab17}, we produce a dominance map, but concentrate on regions closer to the star than in their study.
\begin{figure}
    \centering
    \includegraphics[width=0.9\linewidth]{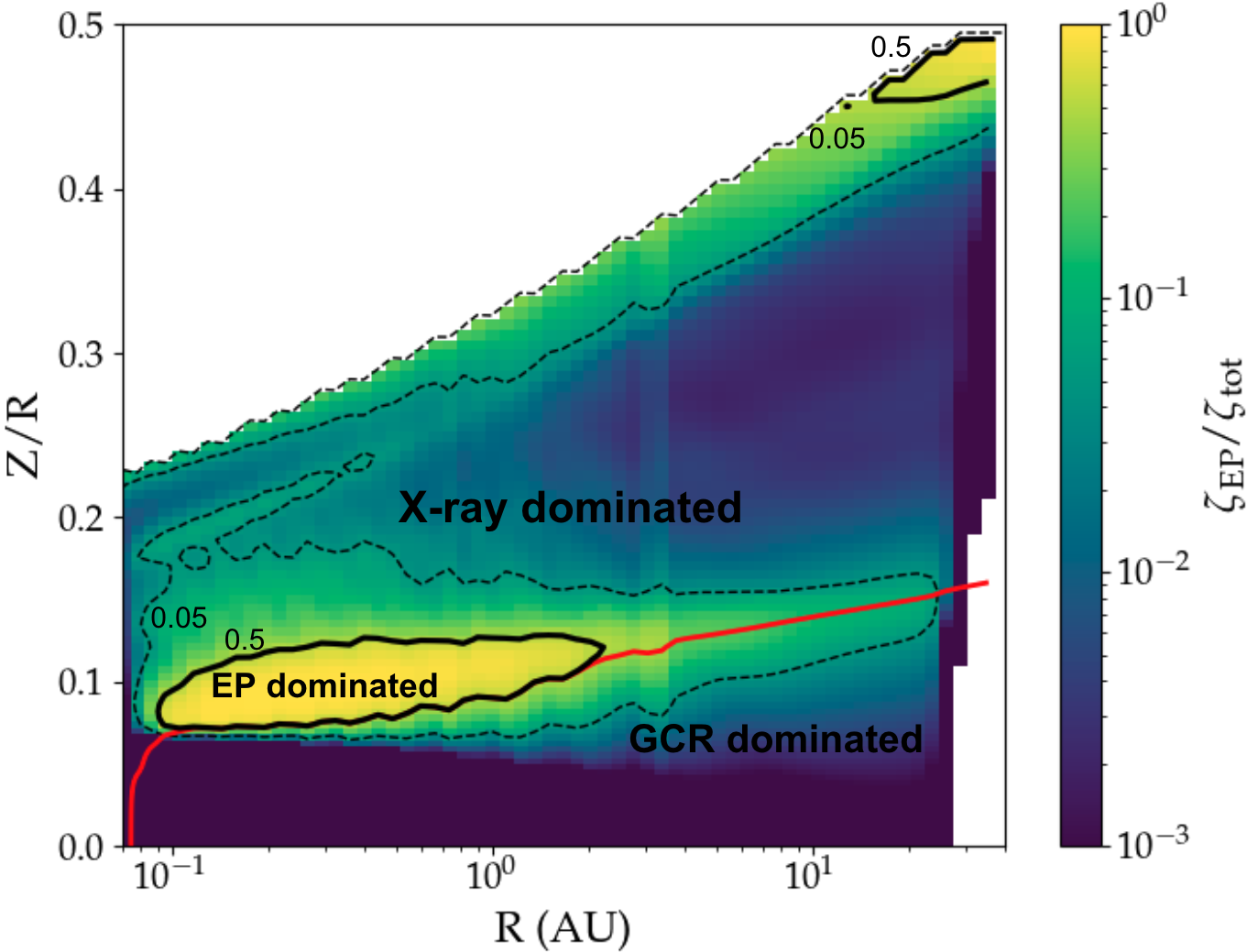}
    \caption{Ionisation rate due to EPs ($\zeta_{\text{EP}}$) compared to the total ionisation rate $\zeta_{\rm tot}= \zeta_{\text{XR}}+\zeta_{\text{GCR}}+\zeta_{\text{EP}}$, for the reference case $\kappa=0.1$, where $\zeta_{\text{XR}}$ and  $\zeta_{\text{GCR}}$ are the ionisation rates due to X-rays and GCRs, respectfully. The solid black contours mark regions where ionisation from EPs dominates over all other ionisation sources, while the dashed black line indicates where the EP ionisation is 5\% of the total ionisation rate. The red contour outlines regions dominated by GCRs. The region in between is dominated by X-ray ionisation.}
    \label{fig:DominantIonisationSourceRedFact10}
\end{figure}

Fig. \ref{fig:DominantIonisationSourceRedFact10} shows the fraction of the total ionisation rate produced by EPs, $\zeta_{\rm EP}/\zeta_{\rm tot}$, for our reference case $\kappa =0.1$, with $\zeta_{\rm tot}=\zeta_{\rm EP}+\zeta_{\rm XR}+\zeta_{\rm GCR}$. The colour scale is logarithmic. Bright yellow marks regions where EPs contribute most. Solid black contours enclose zones where EPs dominate, i.e. $\zeta_{\rm EP}>\zeta_{\rm XR}+\zeta_{\rm GCR}$. The dashed black contour marks where EPs supply a non-negligible share of the ionisation rate, at least 5\% of $\zeta_{\rm tot}$. The red contour outlines regions dominated by GCRs. Fig. \ref{fig:DominantIonisationSourceRedFact10} highlights the efficiency of EP ionisation even when only 10\% of the local viscous dissipation is channelled into non-thermal particles. Even with this modest energy budget, EPs form a clear dominant band in the inner disc, between $0.1-2$ AU in the intermediate to deep layer at $Z/R\sim0.1$. Beyond the strictly dominant zones, large areas exceed the 5\% threshold at distances between $0.1-20$ AU from the star, in the disc atmosphere and in the intermediate to deep layer between $Z/R\sim 0.8-0.15$. These zones of EP dominance are largely determined by the EP energy and propagation depth. In particular, GCRs due to their higher energies, propagate at higher column density and dominate the ionisation close the disc mid-plane. While X-rays dominate in the surface to intermediate region, above the EP-dominated zone. 
The EPs are thus a significant contributor to the ionisation rate over extended regions, relevant for disc chemistry and dynamics.

\begin{figure}
    \centering
    \includegraphics[width=0.9\linewidth]{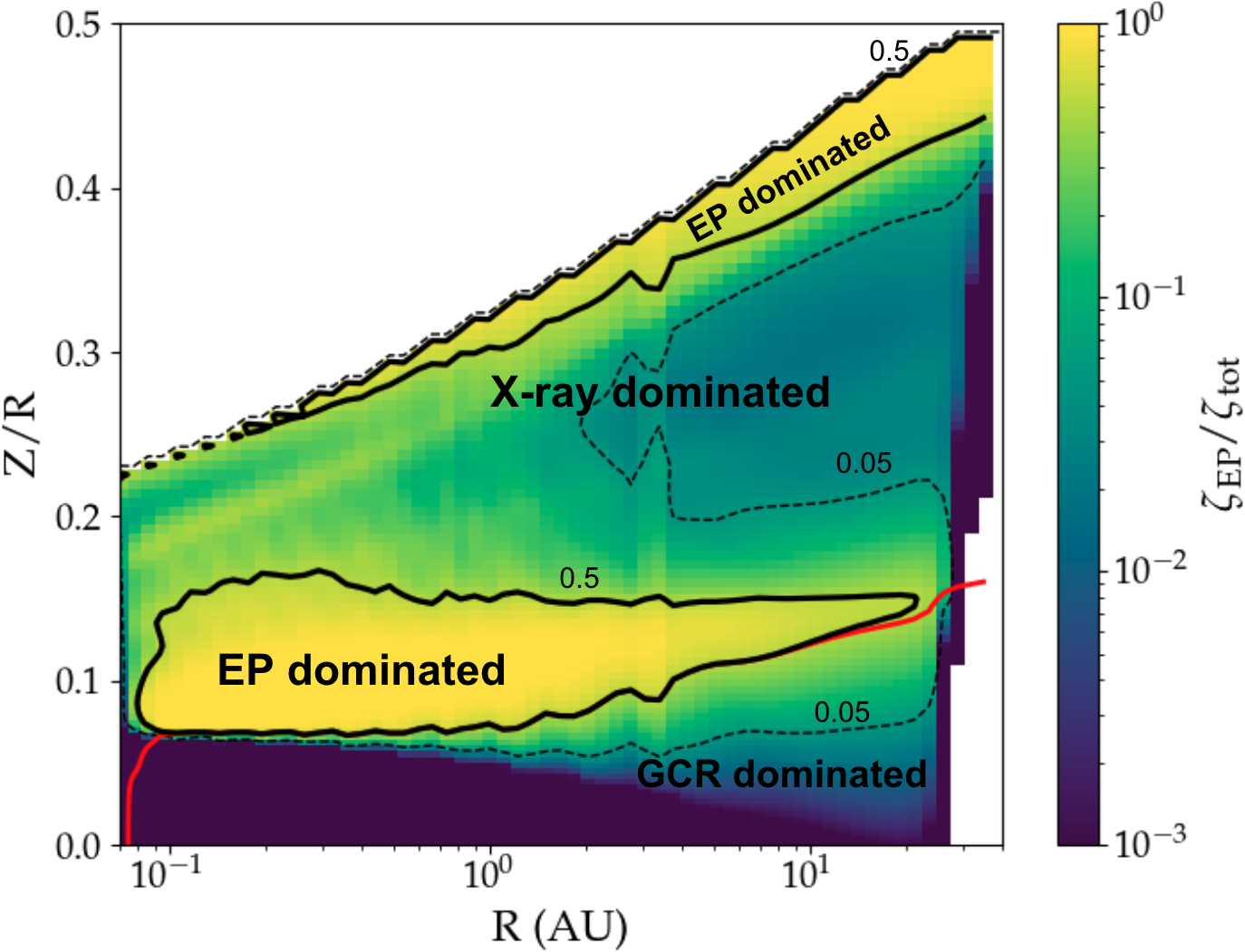}
    \caption{Same as Fig. \ref{fig:DominantIonisationSourceRedFact10}, but for $\kappa=1$.}
    \label{fig:DominantIonisationSource}
\end{figure}
Fig. \ref{fig:DominantIonisationSource} shows the spatial distribution of the EP contribution to the total ionisation rate under the extreme assumption that all local viscous dissipation is channelled to non-thermal particles, i.e. $\kappa=1$. The ionisation map shows that EPs dominate in the disc atmosphere from $\sim 0.2 - 40$ AU, and also in intermediate to deep layers from $\sim 0.08 - 20$ AU. This figure highlights that EPs accelerated by turbulence-induced magnetic reconnection can dominate the ionisation in several regions of the disc, and if not dominant, contribute significantly (more than 5\%) to the total ionisation rate in most of the disc at radii from 0.1 AU to tens of astronomical units.

The $\kappa=1$ assumption is unrealistic, but we use it to illustrate the upper-limit spatial pattern that our model can produce. We note that a similar ionisation state could also occur in systems with higher accretion rates than adopted here. The value $\dot{M}=10^{-8} M_\odot$ yr$^{-1}$ is a conservative choice. A larger $\dot{M}$ increases the viscous dissipation power, and therefore the energy available for EP acceleration. In this sense, the $\kappa=1$ map can be viewed either as an upper bound or as a proxy for cases with $\kappa<1$ but higher $\dot{M}$, which can produce EP-dominated regions of comparable extent and depth. The above results demonstrate that EPs, previously unexplored, may be a major ionisation source in thermochemical models of protoplanetary discs and could be relevant in planetary astrophysics and prebiotic chemistry.

\subsubsection{Energetic thresholds for the ionisation impact of reconnection-accelerated particles} \label{sec:energythreshold}

To test the robustness of our conclusions, we varied the fraction of viscous dissipation power channelled into EP acceleration and examined the resulting ionisation rates. By reducing this fraction $\kappa$, we quantified how sensitive EP-driven ionisation is to the available energy budget and identified when EPs remain a significant source:
\begin{itemize}
    \item When $\kappa < 0.004$, i.e., less than 0.4\% of the local volumetric viscous dissipation is channelled into non-thermal energy, EPs do not dominate the ionisation in any region of the disc. Under these conditions, X-rays and GCRs remain the prevailing sources of ionisation throughout the disc.
    \item For $\kappa < 2.5\times10^{-4}$ (0.025\% efficiency), EPs contribute to less than 5\% of the total ionisation rate ($\zeta_{\rm EP}< 0.05\, \zeta_{\rm tot}$) across the entire disc. In this case, their influence on the ionisation structure, and consequently on the chemistry, magnetic coupling, and observable signatures, is likely negligible.
\end{itemize}
These results show that ionisation by EPs from turbulent magnetic reconnection depends sensitively on $\kappa$. EP-dominated regions appear for efficiencies $\gtrsim0.4\%$, and even a very small conversion of $0.025\%$ yields local contributions at the level of a few percent. Moreover, although subdominant, such localised enhancements are chemically and dynamically relevant. They can modify molecular abundances, shift the ionisation balance that controls magnetic coupling, and imprint distinct line-emission features. Crucially, this behaviour differs from X-rays and GCRs, whose uncertainties (even by factors of a few) tend to affect the entire disc more uniformly. The locality of EP-driven ionisation therefore offers an observational handle. Standard GCR and X-ray rates may reproduce molecular-ion emission across much of a disc, yet still struggle to model the inner disc ionisation. Indeed, forward modelling of DM Tau \citep{2024ApJ...972...88L} finds that models with X-rays and GCR ionisation matching the observations in the outer disc still under-produce inner-disc HCO$^+$, H$^{13}$CO$^+$, and N$_2$H$^+$ compared to observations below 50 AU, indicating an additional, more localised ionisation source in the inner regions. Accordingly, EPs accelerated by magnetic reconnection, even under conservative assumptions, should be included in thermochemical and dynamical disc models. Moreover, targeted inner-disc observations of molecular ions (e.g. HCO$^+$, H$^{13}$CO$^+$, N$_2$H$^+$) can help to discriminate EPs from globally acting X-rays and GCRs via their confined spatial signatures. We also note that this study neglects (i) re-acceleration by MRI-driven turbulence \citep{2021MNRAS.506.1128S} and (ii) other EP acceleration channels \citep{2016ApJ...816...25P}, which could further strengthen or reshape these localised effects.

\subsection{Ionisation volumetric heating rate}
Ionisation not only sets chemistry but is also expected to heat the gas. When EPs ionise H or H$_2$, energy is deposited locally through elastic and inelastic collisions and subsequent chemistry. Following the standard prescription of \citet{Glassgold_2012}, we took per–ionisation heat deposition $Q_{\rm H}=4.3$ eV and $Q_{\rm H_2}=18$ eV, and computed the EP volumetric heating rate as
\begin{equation}
\Gamma_{\rm EP}(R,Z) = \left(n_{\rm H} Q_{\rm H} + 2 n_{\rm H_2} Q_{\rm H_2}\right)\zeta_{\rm EP},
\label{eq:heatingrategeneral}
\end{equation}
where $n_{\rm H}$, $n_{\rm H_2}$ are the local atomic, molecular hydrogen number density, respectively and $\zeta_{\rm EP}$ is the EP ionisation rate per H nucleus, computed in Eq. \eqref{eq:totalionisationrate}.\\
This microphysically-based estimation of the heating term can be directly incorporated into thermochemical models of protoplanetary discs and outflows. Indeed, previous studies have shown that heating at the disc surface significantly affects the launching conditions of photo-evaporative and MHD-driven winds \citep{2000A&A...361.1178C, 2016ApJ...818..152B,2024A&A...686A.287M}. While these studies typically include stellar X-ray and UV heating as primary sources, we now argue that they should consider EPs. 
\begin{figure}
    \centering
    \includegraphics[width=0.9\linewidth]{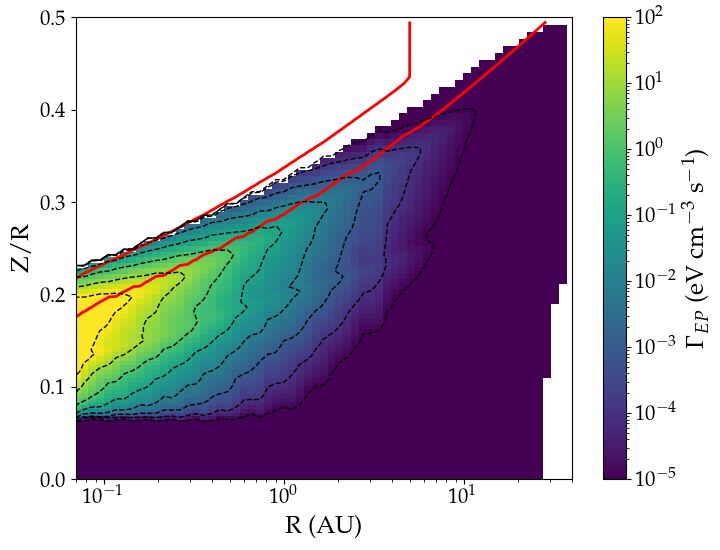}
    \caption{EP volumetric heating rate map for $\kappa=1$. The solid red contour encloses the region where EPs are accelerated. Dashed curves are isocontours of the EP volumetric heating rate. The upper white band marks very low columns ($N<10^{15} \mathrm{cm^{-2}}$), i.e. above the disc atmosphere.}
    \label{fig:EPVolumetricHeatingMap}
\end{figure}

Fig. \ref{fig:EPVolumetricHeatingMap} shows the spatial distribution of the EP volumetric heating rate, $\Gamma_{\mathrm{EP}}$ (eV cm$^{-3}$ s$^{-1}$), assuming the extreme $\kappa =1$ case. Heating is the strongest in the inner disc (within a few astronomical units) and at intermediate heights ($Z/R \sim 0.15-0.3$), where EP flux and gas density combine to maximise energy deposition, and it gradually declines toward larger radii and deeper layers. The solid red contour marks the zone where $E_{\max}>10~\mathrm{MeV}$ (efficient EP acceleration), dashed lines are isocontours, and the white region at the top indicates a very low column density ($N<10^{15} \mathrm{cm^{-2}}$), above the disc atmosphere. These results indicate that EPs accelerated in the atmosphere channel accretion and magnetic energy into heating the gas deeper in the disc.
\begin{figure}
    \centering
    \includegraphics[width=0.9\linewidth]{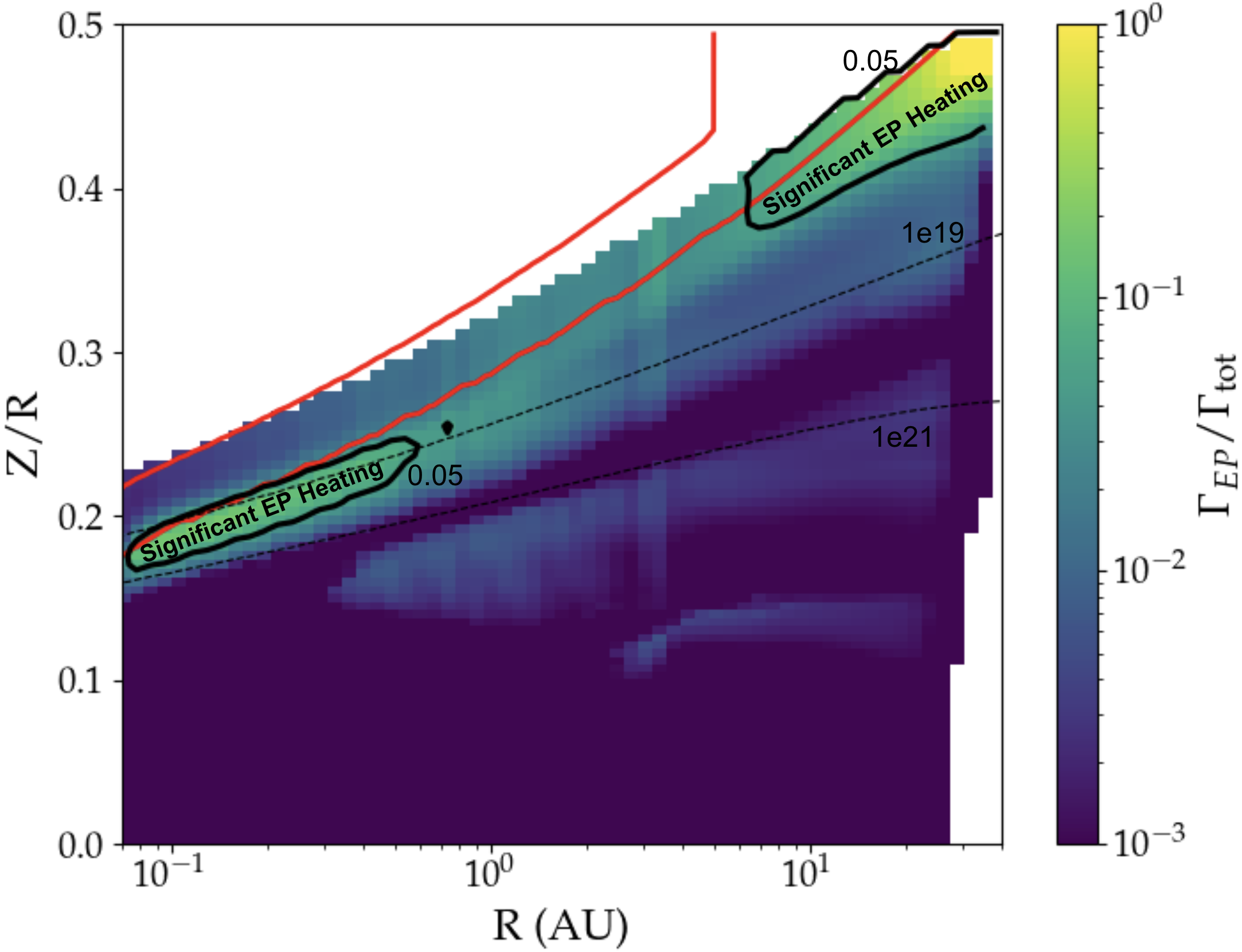}
    \caption{Relative EP heating map for $\kappa=1$. Ratio $\Gamma_{\rm EP}/\Gamma_{\rm tot}$ as a function of radius $R$ (AU) and height $Z/R$; colours show the fraction on a logarithmic scale from $10^{-3}$ to 1. The red contour encloses the zone where EPs are efficiently accelerated ($E_{\max}>10$ MeV). The solid black contour marks where EP heating contributes at least to $5\%$ of the total heating ($\Gamma_{\rm EP}/\Gamma_{\rm tot}=0.05$), which is considered to be a significant contribution. The dashed lines delimit the disc surface between column densities of $10^{19}-10^{21}$ cm$^{-2}$.}
    \label{fig:RelativeEPVolumetricHeatingMap}
\end{figure}
We compared the volumetric heating from EPs, $\Gamma_{\rm EP}$, to the total heating budget, $\Gamma_{\rm tot}$, computed by \prodimo (photoelectric, $\mathrm{H_2^*}$ de-excitation, $\mathrm{H_2}$ formation and dissociation, gas–dust accommodation, X-ray Coulomb, viscous, chemical, etc.). Fig. \ref{fig:RelativeEPVolumetricHeatingMap} maps the ratio $\Gamma_{\rm EP}/\Gamma_{\rm tot}$ throughout the disc. Even in the upper-limit case $\kappa=1$, EP heating is negligible near the mid-plane and generally small elsewhere (mostly $10^{-3}-10^{-2}$), becoming significant only in a narrow region at the disc surface in the inner AU and reaches close to unity values in the disc atmosphere beyond $\sim10$ AU. For more realistic efficiencies, $\kappa\sim0.1$, the EP contribution stays below $\sim1\%$ across the disc, except for a small region at the surface around $R\gtrsim10$ AU where it reaches the few percent level. These fractions are well within typical uncertainties of the total heating budget. Translated into temperature, the impact is expected to be minor compared with other heating sources. Consequently, EP heating appears secondary for thermochemical modelling, and its omission is unlikely to affect global temperature structures. By contrast, the EP ionisation term is the robust, model-relevant effect. It persists at large columns where X-rays are attenuated and can effectively modify chemistry. We therefore emphasise EP ionisation as the key addition, while treating EP heating as optional for extreme or surface-layer wind launching studies.

\section{Discussion}
\label{S:DIS}
In this section we summarise our findings and place them in context. We outline dynamical implications for accretion and wind launching, before closing with the main limitations of our framework and priorities for future work.
\subsection{Turbulence-induced ionisation in discs: Interpretation, and thresholds}

As we have shown, EPs accelerated in situ by turbulence-induced magnetic reconnection in disc atmospheric layers can substantially reshape the ionisation structure of protoplanetary discs. When the non-thermal power is a fraction $\kappa \gtrsim10^{-2}$, EP ionisation overtakes stellar X-rays and GCRs across extended regions, inner radii and surface to intermediate layers, while even $\kappa\sim10^{-3}$ yields contributions of a few percent that are still chemically relevant. Because EPs penetrate to large column densities, they maintain significant ionisation where X-rays are attenuated, providing an additional heating channel at depth.

This behaviour follows naturally from expected atmospheric conditions: good magnetic coupling in far UV/X-ray–ionised layers, sustained MRI turbulence with $\varv_{\rm turb} \sim 0.1 - 0.3 c_s$ and sub$-H$ coherence, and fast turbulent reconnection with $ V_r \sim V_A M_A^2$. In such environments, the turbulent cascade continually forms thin current sheets that release magnetic energy and inject supra-thermal particles. Modelling the acceleration with a Fermi-like prescription yields a steady non-thermal tail with $p\simeq2.5$, bounded by cut-offs set by geometry and timescales, and normalised by the local energy budget via $\kappa$. This adds a disc-internal heating and ionisation channel that complements UV, X-rays, and GCRs, and remains effective at depths where external ionisers fade, with implications for both chemistry and non-ideal MHD.\\
Practically, for the T Tauri system considered here ($\dot{M}=10^{-8} M_\odot \mathrm{yr}^{-1}$ and $L_X=10^{30}$ erg s$^{-1}$) three regimes emerge: 
\begin{itemize}
    \item $\kappa \gtrsim 10^{-2}$: The EPs dominated ionisation in the inner radii in the intermediate/deep layer and in the atmosphere. 
    
    \item $10^{-3} \lesssim \kappa \lesssim 10^{-2}$: The EPs are subdominant yet chemically significant (at the level of a few percent) in extended regions and can still matter for specific chemical tracers and non-ideal MHD terms.
    
    \item $\kappa \lesssim 10^{-3}$: The EP effects on discs can be considered insignificant.  
\end{itemize} 
Here $\kappa$ encapsulates the partition of released magnetic energy into heat versus particles, and electron–proton sharing, and is likely height- and radius-dependent. Here we set $\kappa \approx 0.1$, by analogy with the canonical efficiency with which supernova-remnant shocks are thought to channel kinetic energy into GCR, in a future work, a disc-specific, first-principles estimate of $\kappa$ will be developed.

The maps are most sensitive to three ingredients: (i) the available non-thermal power, $\kappa$; (ii) the ability to reach mega-electronvolt energies (controlled by $\alpha_{\rm acc}$, $\tau_{\rm esc}$, $\tau_{\rm inj}$, and geometry $L_r/H, M_A$); and (iii) local gas thermodynamics through the viscous parameter, $\alpha_{\rm eff}$, fixing the turbulent speed. 

\subsection{Implications for chemistry and observables}

We expect reconnection-accelerated EPs to measurably alter the disc chemistry because their additional ionisation directly modifies the ion–molecule network and the electron fraction in the layers that they penetrate. As for X-rays and GCRs, EP ionisation of H$_2$ initiates the H$_2^+\rightarrow$ H$_3^+$ pathways, enhancing the abundance of classical ionisation tracers (e.g. HCO$^+$, N$_2$H$^+$) and shifting their emitting layers. Cascades of secondary electrons further generate a CR-like internal UV field \citep[Prasad–Tarafdar effect][]{1983ApJ...267..603P,2024A&A...682A.131P}, boosting photodissociation, ionisation and grain charging at high column densities. Because the ionisation rate drives the abundance of nitriles and hydrocarbons (e.g. HCN, C$_2$H$_2$), introducing $\zeta_{\rm EP}(R,Z)$ where X-rays are attenuated will raise n$_e$, push ion–molecule chemistry to larger column density, and alter abundances and lines of these tracers, yielding observable signatures. To quantify these effects, the second paper in this series of articles, will post-process our disc backgrounds with \prodimo, adding to the standard UV/X-ray/CR sources, the EP terms $\zeta_{\rm EP}(R,Z)$ and $\Gamma_{\rm EP}(R,Z)$. We will explore a grid in non-thermal power fraction $\kappa$, spectral slope $p$, and acceleration geometry; solve to thermochemical equilibrium with and without $\zeta_{\rm EP}$; and map changes in gas temperature, electron fraction, and the layers where EPs overtake X-rays/GCRs. We will quantify EP-driven changes in classic ionisation tracers (HCO$^+$, N$_2$H$^+$) and, importantly, in mid-IR prebiotic and hydrocarbon species (e.g. HCN, HNC, C$_2$H$_2$) that originate in the warm surface layers where EPs effect is strong, exactly the regions probed by JWST/MIRI. We will generate synthetic MIRI diagnostics (e.g. band fluxes, ratios, emitting heights, line-to-continuum ratio) to assess how EPs imprint on those spectra and how to observe them. The outcome will be abundance and temperature maps and practical discriminants that isolate EP-driven ionisation from X-ray/CR scenarios and delineate the ($\kappa,p$) domains where EP effects remain observable.

\subsection{Dynamical implications for accretion and winds}

Reconnection-accelerated EPs ionise and raise the electron fraction, $n_e$, in layers where stellar X-rays are already attenuated. In the simple gas-phase recombination balance, $n_e,  \sim  \sqrt{\zeta /(\beta_{\rm rec} n_{\rm H})}$, where $\beta_{\rm rec}$ is the reconnection rate \citep{fromang2002ionization}, so boosting the ionisation rate by a factor, $f$, increases $n_e$ by$ \sim f^{1/2}$. The EP contributions can therefore significantly increase $n_e$, with two immediate consequences. 

First, on accretion and MRI coupling activation, larger $n_e$ reduces Ohmic resistivity $\eta_{\rm O} \propto 1/n_e$ and, by increasing the ion density, reduces ambipolar diffusivity, $\eta_{\rm A} \propto 1/\rho_i$. The corresponding Elsasser numbers, $\Lambda_{\rm O}=v_A^2/(\eta_{\rm O}\Omega)$ and $Am=v_A^2/(\eta_{\rm A}\Omega)$, rise. Layers that were marginal (almost active, $\Lambda_{\rm O} \sim 1, Am \sim 1 )$ can cross MRI thresholds, shrinking dead zones and pushing the magnetically active region to larger column densities, which will increase the local accretion rate. This creates a bootstrap process: MRI produces turbulent reconnection that accelerates EP, which enhances ionisation triggering further MRI activation. In this picture, EPs produced in an MRI-active region may irradiate adjacent, marginally coupled layers, increasing the EP acceleration region size. A saturation arises from the $\beta$–dependence of EP acceleration. Turbulent-reconnection acceleration is most efficient in low-$\beta$ regions; as the MRI-active layer expands and encroaches on higher-$\beta$ gas, the reconnection-driven EP production drops. Once the front of the MRI-active zone reaches these high-$\beta$ layers, it can no longer sustain a supra-thermal EP population able to ionise, so the outward growth of the acceleration region naturally stalls, quenching the bootstrap loop. A related pathway may operate in regions where the field is still sufficiently frozen-in to wind up. Reversals between toroidal field loops can generate thin current sheets even when MRI modes are suppressed. These sheets can undergo bursts of turbulent reconnection, producing rapid EP acceleration and temporary spikes in ionisation, potentially activating MRI and EP acceleration locally and intermittently.

Second, on wind launching and magneto-thermal coupling, magneto-thermal winds require gas that is well coupled to a reasonably coherent large-scale field at the launch base. Disc layers where winds are anchored are expected to be turbulent and only partially ionised, which lowers conductivities and suppresses efficient wind launching solutions. By adding $\zeta_{\rm EP}(R,Z)$ where X-rays fade, EPs can thicken the well-coupled layer and, through their volumetric heating, $\Gamma_{\rm EP}$, aid mass loading. This does not guarantee large-scale field coherence, but it may broaden the parameter space in which a steady magneto-thermal wind can be maintained, shifting the launch region and altering mass-loading and torque.

In a future work, we will post-process this additional EP source on a ProDiMo disc structure to clarify the importance and spatial extent of this bootstrap process and whether this mechanism can patch or reduce the dead zone. However, only dedicated non-ideal MHD simulations, including turbulent reconnection and EP transport self-consistently, can properly assess the impact of this mechanism. Including $\zeta_{\rm EP}(R,Z)$ and $\Gamma_{\rm EP}(R,Z)$ in conductivity and energy modules will modify vertical profiles of $\eta_{\rm O}$, $\eta_{\rm A}$ (and Hall where relevant), the MRI-active layer thickness, and the wind-launch region. Because EPs penetrate to a larger column density than X-rays, their impact is concentrated exactly where simulations currently struggle to maintain adequate coupling between the disc material and magnetic field. While EPs will not by themselves 'solve' the wind-launching problem, they provide an additional, physically motivated ionisation and heating channel that can tip marginal layers into the coupled regime, with measurable consequences for both accretion and outflows.

\subsection{Limitations}

Our framework entails several simplifying assumptions. First, on the acceleration model. We adopted a Fermi-like gain rate, $\dot{E}=\alpha_{\rm acc} E$, and connected $\alpha_{\rm acc}$, $\tau_{\rm esc}$, and $\tau_{\rm inj}$ to a turbulent reconnection geometry. While these estimates are consistent with reconnection-in-turbulence theory and numerical trends, several efficiencies remain poorly constrained and depend on disc height and radius: (i) the partition of released magnetic energy into particles versus heat; (ii) the electron–proton sharing, for which future work will include injection and acceleration efficiencies, $\eta_e,\eta_p$; (iii) the fraction of turbulent power that feeds reconnection, set by the statistics and filling factor of current sheets; and (iv) the role of guide-field geometry and plasma, $\beta_p$, in regulating $\alpha_{\rm acc},\ \tau_{\rm esc}$ and pitch-angle scattering. To make these dependencies explicit and to avoid over-constraining uncertain microphysics, we normalised the non-thermal component by the local energy budget through $\kappa$. The phenomenological parameter $\kappa$ thus absorbs uncertainties in the reconnection, particle and heat partition, and species sharing, allowing us to assess how the ionisation maps respond to realistic variations in the acceleration efficiency, while remaining grounded in the turbulent–reconnection geometry. A physically grounded estimation of $\kappa$ will be provided in a future publication.
    
Second, on particle transport, we modelled EP propagation using the CSDA, neglecting pitch–angle scattering. CSDA is reliable up to column densities of order $\sim 10^{25} \mathrm{cm^{-2}}$; beyond this depth (i.e. for protons above the pion–production threshold) a transport treatment including spatial diffusion and discrete losses is required \citep{Padovani18}. In this set-up we also approximate the magnetic field as locally vertical, omitting its toroidal and poloidal components. A magnetic field configuration, combining toroidal shear and poloidal lines, would induce helical guiding-centre motion, thereby modifying path lengths and redistributing EP flux laterally. Magnetic field wandering would also lengthen particle path–lengths and generally reduce penetration depths (tending to shift EP–dominated regions upwards in $Z/R$), and may inject particles at higher radii. In addition, field–line channelling could focus fluxes in preferred directions; capturing these effects would demand coupling transport to self-consistent MHD fields. Furthermore, in turbulent flows, non-thermal particles can further gain energy by stochastic acceleration in randomly moving MRI-generated magnetic perturbations  \citep{2021MNRAS.506.1128S}. In this situation, non-thermal particles can also get (re)accelerated by Fermi second order process and shear acceleration. The particle maximum momentum, $p_{\rm max}$, scales as the ratio of the local sound speed to light speed, $ c_s/c$. Because of the small ratio $c_s/c$, it is clear that the acceleration effect is rather modest. Quantifying such re-acceleration is deferred to future work.

Third, on time dependence and intermittency, we computed the non-thermal distribution at the end of a current-sheet lifetime and then averaged over acceleration sites on a stationary disc background. The accretion rate used is likewise steady and conservative ($\dot{M}=10^{-8} M_\odot \mathrm{yr}^{-1}$), i.e. it does not include episodic enhancements. In reality, accretion, reconnection, and turbulence are intermittent, and young discs show non-stationary accretion with frequent bursts. During such bursts, a larger fraction of the (temporarily increased) accretion power can be channelled into EPs, raising the ionisation and heating rates ($\zeta_{\rm EP}$, $\Gamma_{\rm EP}$), pushing EP influence deeper in the disc and driving time-variable chemical responses. Incorporating burst statistics and fully time-dependent transport and thermochemistry is a natural next step, beyond the steady-state framework adopted here.

As a perspective three next steps are appearing. (1) Thermo-chemical feedback and observables: post-process with \textsc{ProDiMo} including $\zeta_{\rm EP}(R,Z)$ and $\Gamma_{\rm EP}(R,Z)$ to predict changes in abundances, temperatures, and synthetic observational ALMA/JWST diagnostics, and to identify discriminants between EP-, X-ray-, and CR-dominated regimes. (2) Calibration of efficiencies: constrain $\alpha_{\rm acc}$, $\tau_{\rm esc}$, $\tau_{\rm inj}$ and particle and heat partition from stratified-box simulations that measure reconnection statistics, and confront these with observational diagnostics.  (3) Transport: couple EP propagation to 3D MHD fields to include field-line diffusion, and stochastic re-acceleration, and extend beyond the CSDA limit.

\section{Conclusion}
\label{sec:conclusions}

We investigated the ionisation produced by EPs accelerated in situ by turbulence-induced magnetic reconnection in the atmospheres of protoplanetary discs. Building on a Fermi-like prescription for the energy gain and linking the acceleration, escape and injection timescales to a turbulent–reconnection geometry \citep{2023ApJ...942...21X}, we derived a non-thermal particle distribution with a steady-state spectral index $p=2.5$, an energy range bounded by a turbulence– and geometry–set cut-off, and a normalisation constrained by the fraction of accretion energy channelled to EPs through the parameter $\kappa$. Using a \textsc{ProDiMo} disc background, we computed EP propagation and the resulting ionisation and heating rate spatial distribution.\\
Our main conclusions are the following.
\begin{enumerate}
\item A new leading ionisation source.
For $\kappa \gtrsim 10^{-2}$, EPs dominate the ionisation budget over stellar X-rays and GCRs in the inner disc (sub-astronomical units to a few astronomical units) at normalised height $Z/R \approx 0.1$ and in the disc atmosphere out to radii of a few tens of astronomical units (see Figs.~\ref{fig:DominantIonisationSource}, \ref{fig:DominantIonisationSourceRedFact10}). Even for $\kappa \sim 10^{-3}$, EPs contribute at the level of a few percent locally – enough to affect chemistry and magnetic coupling. Below $10^{-3}$, they are minor.
\item Deep penetration. Compared to X-rays, EPs maintain stronger ionisation at large column densities, modifying the ionisation structure in layers that are otherwise shielded from radiation. 
\item Heating. EP-driven heating is typically modest compared to the total heating budget. For $\kappa\lesssim0.1$, $\Gamma_{\rm EP}$ stays below $\sim\!1\%$ almost everywhere, reaching the percent level only in a narrow atmospheric band at $R\!\gtrsim\!10$ AU. Ionisation — not heating — is the primary impact of EPs in our framework.
\item Observability and locality as a test.
Because EPs are generated locally, they imprint spatially confined ionisation patterns distinct from globally distributed GCRs or stellar X-rays. This locality will provide testable signatures in emitting position of ion tracers (HCO$^+$, N$_2$H$^+$) and mid-IR species (HCN, C$_2$H$_2$) in the same layers that JWST (especially MIRI) probes. Comparing these diagnostics against our maps could offer a way to test our model.
\end{enumerate}
Consequently, EPs from turbulent reconnection should be incorporated into thermochemical and dynamical disc models and for the interpretation of ionisation tracer observations.

In summary, turbulence-induced magnetic reconnection in disc atmospheres provides a physically motivated, in situ source of EPs that can reshape the ionisation structure of protoplanetary discs. Even when energetically subdominant, these particles penetrate and affect layers where other ionisers fade. When reconnection is efficient, they become a leading ionisation channel. Accounting for this component is therefore necessary for more realistic models of disc chemistry, dynamics, and their observational signatures. 
In turn, including this EP component in disc models will matter in the broader astrophysical context. In astrobiology, ionisation helps building (and destructing) complex organic molecules, sets where prebiotic species can form, and affects the interpretation of possible biosignatures in young planetary systems. More broadly, the same turbulence–reconnection framework can be adapted to accretion onto compact objects, where it may likewise contribute to the accretion–ejection dynamics.

\begin{acknowledgements}
V.B. and M.P. acknowledge the INAF grant 2023 MERCATOR (``MultiwavelEngth signatuRes of Cosmic rAys in sTar-fOrming Regions'');
M.P. acknowledges the INAF grant 2024 ENERGIA (``ExploriNg low-Energy cosmic Rays throuGh theoretical InvestigAtions at INAF''). CHR acknowledge the support of the Deutsche Forschungsgemeinschaft (DFG, German Research Foundation) Research Unit ``Transition discs'' - 325594231. CHR is grateful for support from the Max Planck Society.
C.S aknowledge support for this work by the "Action Thématiques" "Physique Stellaire" (ATPS), "Phénomènes Extrêmes et Multi-messagers" (AT-PEM) and "Physique Chimie du Milieu Interstellaire (AT-PCMI) of CNRS/INSU PN Astro with INP and IN2P3, co-funded by CEA and CNES". AM thanks the support of IN2P3 through the INTERCOS master project. The authors thank D. Galli, G. Lesur and D.
Rodgers-Lee for fruitful discussions.
\end{acknowledgements}

\bibliographystyle{aa}
\bibliography{biblio}

\appendix
\section{Turbulent viscosity model}\label{annex:Viscosity}
As already stated, in order to develop, MRI requires a sufficiently high degree of ionisation for the gas to interact with the magnetic field. The instability is characterised by the viscous parameter $\alpha_{\rm eff}$. To compute the viscous $\alpha_{\rm eff}$ parameter, we rely on the model proposed by \citet{2019A&A...632A..44T}, which introduces an empirical non-ideal MHD MRI-driven equation for $\alpha_{\rm eff}$ suitable for physico-chemical protoplanetary disc codes such as \prodimo. The adopted expression for $\alpha_{\rm eff}$ is given by 
\begin{equation}
    \alpha_{\rm eff} = \left(\frac{2}{\beta_{\rm mag}}\right)^{1/2} \rm min(1,\Lambda_{\text{Ohm}}) \left[\left(\frac{50}{Am^{1.2}}\right)^2+\left(\frac{8}{Am^{0.8}}+1\right)^2\right]^{-1/2}
    ,
    \label{eq:effectiveviscosityThi}
\end{equation}

if $\sqrt{\beta_{p}} \Lambda_{\text{Ohm}} > 1$ and $\alpha_{\rm eff} \approx 0$ otherwise. 

$Am$ represents the frequency at which neutral particles collides with ions normalised to the Keplerian orbital frequency, 

\begin{equation}
Am \equiv \frac{\nu_{\rm in}}{\Omega_K} = \frac{\beta_{\rm in} n_{\rm charge}}{\Omega_K}= \frac{x_e \beta_{\rm in} n_{\rm tot}}{\Omega_K} ,
\end{equation}

where $n_{\rm charge} = x_{E}n_{\rm tot}$ is the total number density of charged species and $\beta_{\rm in} = 2 \times 10^{-9}   \rm cm^3 s^{-1}$ is the collisional rate coefficient for ions to distribute their momentum to neutrals. Thus,
\begin{equation}
    Am\approx 10^4\left(\frac{x_e}{10^{-4}}\right)\left(\frac{n_n}{10^{10}~\rm cm^{-3}} \right) \left(\frac{R}{1 \rm{AU} } \right)^{3/2} \ .
    \label{eq:ambipolarElsasser}
\end{equation}

The Ohmic diffusivity is quantified by the dimensionless Elsasser number $\Lambda_{\text{Ohm}}$, defined as the ratio of the Lorentz force to the Coriolis force,
\[
\Lambda_{\text{Ohm}} \equiv \frac{B_z^2}{4\pi \rho \eta_O \Omega_K} \equiv \frac{v_A^2}{\eta_O \Omega_K} \equiv \left( \frac{4\pi \sigma_O}{\Omega_K} \right) \left( \frac{v_A}{c} \right)^2,
\]
where $B_z$ is the vertical component of the magnetic field, $v_A$ is the Alfvén speed and $\rho$ is the mass density of the plasma and the Ohmic resistivity is given by $\eta_O = \frac{c^2}{4\pi \sigma_O}$. \citet{2019A&A...632A..44T} derived an approximation for the Ohmic Elsasser number in disc regions relevant to our study where, $\sigma_O \approx \sigma_{e,O}$ and $x_e > 10^{-13}$:
\[
\Lambda_{\text{Ohm}} \approx 1 \left( \frac{T}{100~\rm K} \right)^{1/2} \left( \frac{10^4}{\beta_{p}} \right) \left( \frac{R}{1 \text{ AU }} \right)^{3/2} \left( \frac{x_e}{10^{-9}} \right),
\]
The magnetic term $\beta_{p}(R,Z)$ is the ratio of the thermal pressure to the magnetic pressure, given by \citep{2019A&A...632A..44T},
\begin{equation}
    \beta_{p}(R,Z) = \beta_{\text{mid}}  \frac{P_{\text{th}}(R,Z)}{P_{\text{th}}(R, 0)} = \beta_{\text{mid}} \frac{n(R,Z)T(R,Z)}{n(R,0) T(R,0)},
    \label{eq:defplasmabeta}
\end{equation}
where $\beta_{\text{mid}}= \beta_{p}(R,0)$ is the value of $\beta_{p}$ on the mid-plane. $\beta_{\text{mid}}$ is assumed to be independent of the radius so that it can be determined from the thermal structure and chemical abundances of \prodimo. 
The parameter $\beta_{\text{mid}}$ is typically in the range $10^4-10^6$. We take $\beta_{\text{mid}}=10^4$ as reference value.

Using the $\beta_p$ and $\alpha_{\rm eff}$ fields derived above, we map the turbulence/reconnection parameters across the entire \textsc{ProDiMo} grid. In particular, the local Alfvén speed is
\begin{equation}
    V_A = c_s \sqrt{\frac{2}{\beta_p}},
\end{equation}
and the Alfvénic Mach number is
\begin{equation}
    M_A \equiv \frac{\varv_{\rm turb}}{V_A}
\end{equation}
where $c_s$ is the local sound speed and $\varv_{\rm turb}$ the turbulent velocity computed from Eq. \ref{eq:turbulentscale}.
\end{document}